\documentclass[pre,twocolumn,showpacs,superscriptaddress]{revtex4}

\usepackage{amsmath}
\usepackage{graphicx}
\usepackage{color}

\usepackage{mathrsfs}
\usepackage{amssymb}
\usepackage{subfigure}
\usepackage[british]{babel}
\begin{document}

%%%%%%%%%%%%%%%%%%%%%%%%%%%%%%%%%%%%%%%
\newcommand{\comment}[1]{{\bf (*** #1 ***)}}
\newcommand{\changeJP}[1]{{\bf **JNP: #1 ***}}
%%%%%%%%%%%%%%%%%%%%%%%%%%%%%%%%%%%%%%%

\date{\today}
%\preprint{\today.}

\title{Bubble merging in breathing DNA as a vicious walker problem in
opposite potentials}

\author{Jonas Nyvold Pedersen}
\affiliation{Department of Mathematical Physics, Lund University,
Box 118, 22100 Lund, Sweden}
\author{Mikael Sonne Hansen}
\affiliation{Department of Mathematics, Technical University of
Denmark, Bldg. 303S, Matematiktorvet, 2800 Kongens Lyngby, Denmark}
\author{Tom{\'a}{\v s} Novotn{\'y}}
\affiliation{Department of Condensed Matter Physics, Faculty of
Mathematics and Physics, Charles University, Ke Karlovu 5, 121 16
Prague, Czech Republic and Nano-Science Center, University of
Copenhagen, Universitets\-par\-ken 5, 2100 Copenhagen, Denmark}
\author{Tobias Ambj{\"o}rnsson}
\affiliation{Department of Chemistry, Massachusetts Institute of
Technology, 77 Massachusetts Avenue, Cambridge, MA 02139,
USA}\altaffiliation{Present address: Department of Theoretical
Physics, Lund University, S\"{o}lvegatan 14A, 223 62 Lund, Sweden.}
\author{Ralf Metzler}
\affiliation{Department of Physics, Technical University of Munich,
James Franck Stra{\ss}e, D-85747 Garching, Germany}
\email{metz@ph.tum.de.}


\begin{abstract}
We investigate the coalescence of two DNA-bubbles initially located
at weak domains and separated by a more stable barrier region in a
designed construct of double-stranded DNA. In a continuum
Fokker-Planck approach, the characteristic time for bubble
coalescence and the corresponding distribution are derived, as well
as the distribution of coalescence positions along the barrier.
Below the melting temperature, we find a Kramers-type barrier
crossing behavior, while at high temperatures, the bubble corners
perform drift-diffusion towards coalescence. In the calculations, we
map the bubble dynamics on the problem of two vicious walkers in
opposite potentials. We also present a discrete master equation
approach to the bubble coalescence problem. Numerical evaluation and
stochastic simulation of the master equation show excellent
agreement with the results from the continuum approach. Given that
the coalesced state is thermodynamically stabilized against a state
where only one or a few base pairs of the barrier region are
re-established, it appears likely that this type of setup could be
useful for the quantitative investigation of thermodynamic DNA
stability data as well as the rate constants involved in the
unzipping and zipping dynamics of DNA, in single molecule
fluorescence experiments.
\end{abstract}

\maketitle

\section{Introduction}

Within a broad range of salt and temperature conditions, the
Watson-Crick double-helix \cite{watson} is the equilibrium structure
of DNA. This thermodynamic stability is effected by hydrogen-bonding
between paired bases and by base-stacking between nearest neighbor
pairs of base pairs
\cite{watson,cantor,kornberg,kornberg1,frank,delcourt,krueger,santalucia}.
By an increase of the temperature or by variation of the pH-value
(titration with acid or alkali) double-stranded DNA progressively
denatures, yielding regions of single-stranded DNA, until the
double-strand is fully molten. This is the helix-coil transition
\cite{poland, peyrard_np}. The melting temperature $T_m$ is defined
as the temperature at which half of the DNA molecule has undergone
denaturation \cite{poland,frank,guttmann,wartell}. Typically, the
denaturation starts in regions rich in the weaker
A\emph{denine}-T\emph{hymine\/} base-pairs, and subsequently moves
to zones of increasing G\emph{uanine}-C\emph{ytosine\/} content. The
occurrence of zones of different stability within the genome was
shown to be relevant when separating coding from non-coding regions
\cite{carlon,yeramian}.

However, already at room temperature thermal fluctuations cause rare
opening events of small intermittent denaturation zones in the
double-helix \cite{gueron}. These \emph{DNA bubbles\/} consist of
flexible single-stranded DNA, and their size fluctuates by step-wise
zipping and unzipping of the base pairs (bps) at the zipper forks,
where the bubble connects to the intact double-strand. Initiation of
a bubble in a stretch of intact double-strand requires the crossing
of a free energy barrier $\Delta G_{\rm bubble}$ of some 8 to 12
$k_BT$ at physiological temperature, corresponding to a Boltzmann
factor, often referred to as the cooperativity factor,
$\sigma_0=\exp(-\Delta G_{\rm bubble}/k_BT) \sim 10^{-5}\ldots
10^{-3}$. Once formed below the melting temperature $T_m$, a bubble
will eventually zip close. Above $T_m$, a bubble will preferentially
stay open and, if unconstrained, grow in size until it merges with
other denaturation bubbles, eventually leading to full denaturation
of the double-helix. Constraints against such full unzipping could,
for instance, be the build-up of twist in smaller DNA-rings or the
chemical connection of the two strands by short bulge-loops, compare
Ref.~\cite{altan}.

Biologically, the physical conformations of DNA molecules are
considered of increasing relevance for its function, see, for
instance, the review \cite{ctn} and references therein. In
particular, the existence of intermittent (though infrequent)
bubble domains is important, as the opening up of the Watson-Crick
base pairs by breaking of the hydrogen bonds between complementary
bases disrupts the helical stack. The flipping out of the ordered
stack of the unpaired bases allows the binding of specific
chemicals or proteins, that otherwise would not be able to access
the reactive sites of the bases
\cite{gueron,poland,krueger,frank}. In fact, there exists a
competition of time scales between the opening/closing dynamics of
DNA-bubbles and the binding kinetics of selectively
single-stranded DNA binding proteins \cite{pant,pant1,tobiasrc,somepanwill}.
That the chemical potential of the single-stranded binding proteins does not
lead to full denaturation of the DNA is due to the slow binding of the proteins
when compared to the bubble dynamics \cite{pant,pant1,tobiasrc,somepanwill}.
It is also believed that
DNA-breathing assists in the transcription initiation process
\cite{tobiasprl,tobiasbj,choi,kalosakas}. The quantitative knowledge of
the denaturation dynamics as well as energetics is imperative to a better
understanding of genomic biochemical processes.

DNA-breathing has been modeled extensively in terms of the
Peyrard-Bishop model, that is based on a set of Langevin equations
for the base-base distance in a base pair; the effective attraction
between the bases is represented by model potentials
\cite{peyrard,dauxois,peyrard_cm,campa,peyrard_nonl}. Alternatively,
DNA-breathing can be considered as a random walk process in the free
energy landscape of the Poland-Scheraga model of DNA denaturation,
as the number of broken base pairs turns out to be the slow variable
of the process \cite{altan}. In continuum form, this approach to DNA
bubble dynamics has been described in terms of a Fokker-Planck
equation \cite{hame,hans,kafri}. A discrete description, in which
the coordinate of the random walker corresponds to a specific base
pair, was suggested in Refs.~\cite{bicout,JPC,tobiasrc,tobiasprl},
and the corresponding stochastic simulations analysis of DNA
breathing was introduced in Ref.~\cite{suman}. The influence of a
random energy landscape on bubble localization and dynamics was
studied in Refs.~\cite{hwa,kafri,jeon}, while a framework to include
an arbitrary given sequence of base pairs was developed in
Refs.~\cite{tobiasprl,tobiasbj,tobiaspre}. Endowed with the
sensitivity of their dynamics, DNA constructs were proposed as
nanosensors \cite{tobiasjpc,tobiasctn}. We note that the formulation
in terms of the gradient of the Poland-Scheraga free energy allows
one to explicitly introduce all necessary independent stacking
parameters based on the study \cite{krueger}, see also the
discussion in Ref.~\cite{tobiasbj}. Measuring the dynamics of DNA
bubbles also provides information on the magnitude of the critical
exponent $c$ representing the entropy loss factor of a closed
polymer loop, deciding on the order of the denaturation transition
\cite{poland1,fisher1,poland,kafri1,comment,haochme,blossey}, by
influencing the temporal survival probability of bubbles
\cite{hans,kafri}.

\begin{figure*}
\scalebox{1.0}{\input{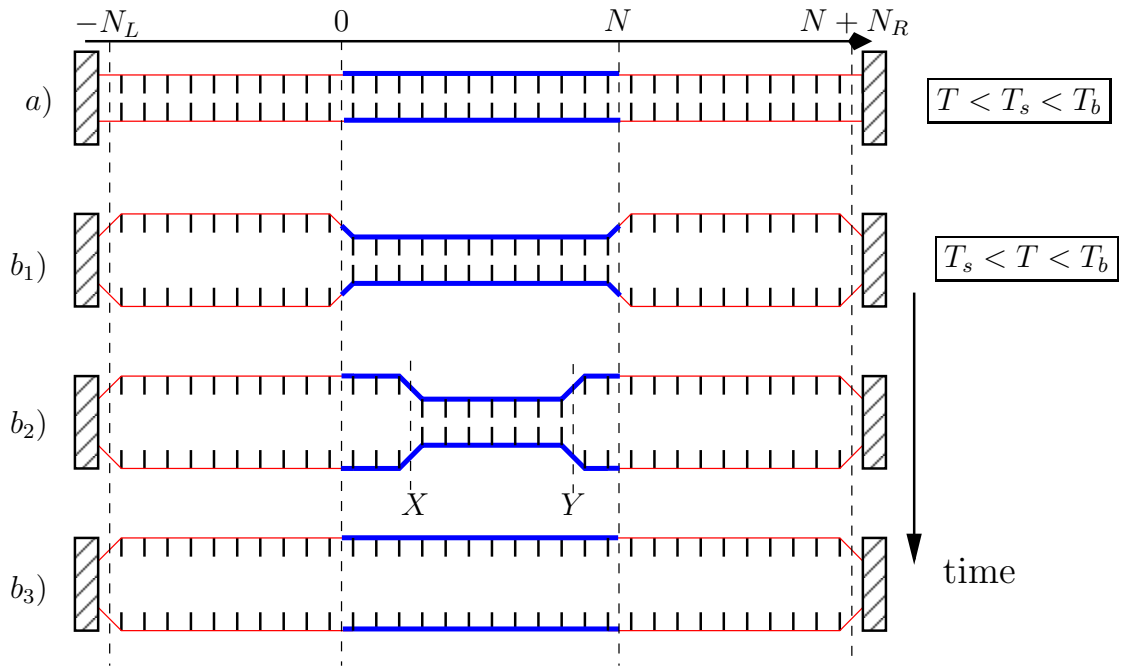}} \caption{ Schematic
picture of the bubble coalescence setup in a designed DNA construct.
It is clamped at both ends and consists of two outer soft zones
(thin red lines) of lengths $N_L,N_R$ bps with melting temperature
$T_s$ and a stronger $N$-bps-long barrier zone (thick blue lines)
with $T_b>T_s$. a) All bps closed ($T<T_s<T_b$). b) Soft zones open
by raising the temperature above $T_s$. b$_1$--b$_3$) Successive
opening of the barrier driven mainly by fluctuations ($T<T_b$) or
drift ($T>T_b$) until coalescence. The discrete coordinates
$X,Y=-N_L,\dots,N+N_R$ are defined as the positions of the {\em
interfaces} between the closed and broken bps. } \label{fig:model}
\end{figure*}

The multi-state nature of \emph{DNA breathing\/} can be monitored in
real time on the single DNA level by fluorescence correlation
techniques \cite{altan}. It has been shown in a quantitative
analysis that the experimentally accessible autocorrelation function
is sensitive to the stacking parameters of DNA
\cite{tobiasprl,tobiasbj}. However, it has not been fully
appreciated to what extent the fluorophore and quencher molecules,
that are attached to the DNA construct in the experiments reported
in Refs.~\cite{altan,olegrev,beacon}, influence the stability of
DNA. Moreover, the zipping rates measured in the single molecule
fluorescence setup differ from those determined in NMR experiments
\cite{gueron,altan}. We here propose and study a complementary setup
for the single molecule fluorescence investigation of DNA breathing,
as shown in Fig.~\ref{fig:model}. In this setup, a short stretch of
DNA, clamped at both ends, is designed such that two soft zones
consisting of weaker AT-bps are separated by a more stable barrier
region rich in GC bps. For simplicity, we assume that both soft
zones and barrier are homopolymers with a bp-dissociation free
energy $\Delta G_s$ and $\Delta G_b$, respectively, and, in
accordance with the experimental findings of reference \cite{altan},
we neglect secondary structure formation in the barrier zone. At
temperatures higher than the melting temperature $T_s$ of the soft
zones, but still lower than the melting temperature $T_b$ of the
barrier region such that two open bubbles are being promoted,
thermal fluctuations will gradually dissociate the barrier, until
the two bubbles coalesce. Note that the melting temperature at 100
mM salt conditions differs by approximately 50 degrees between mixed
(AT/TA)$_n$ and (GC/CG)$_n$ homopolymers respectively
\cite{santalucia,krueger}. This should provide a large enough
temperature interval between hard and soft zones to perform this
type of experiment. In the following we use realistic values for the
simulations. Once coalesced, the free energy corresponding to one
cooperativity factor $\sigma_0\sim 10^{-5}\ldots 10^{-3}$ is
released, stabilizing the coalesced bubble against reclosure of the
barrier. This fact should allow for a meaningful measurement of the
coalescence time in experiment, and therefore provide a new and
sensitive method to measure DNA stability data and base pair zipping
rates. We also study the case when the system is prepared as above
and then $T$ suddenly increased such that $T>T_b>T_s$ so that the
system is driven towards coalescence. In both cases the two
boundaries between bubbles and barrier perform a (biased) random
walk in opposite free energy potentials.

In fact, the study of the bubble coalescence is of interest in its
own right, as we map the random walk of the two zipper forks
separating double-stranded barrier base-pairs from already denatured
single-stranded bubble domains onto a new case of the vicious walker
problem. Namely, we deal with two vicious walkers in \emph{linear\/}
but \emph{opposite\/} potentials. The viciousness condition
corresponds to the fact that when the two zipping forks meet, the
bubbles coalesce, and the dynamics is stopped. While the problem of
a general number of (otherwise noninteracting) vicious walkers in
free space was solved long time ago \cite{fisher} and has been only
relatively recently generalized to the case of motion in a common
potential \cite{bray}, even two walkers in different potentials
cannot be addressed analogously by the straightforward
antisymmetrization procedure. To solve our problem in the continuum
limit, we use a trick of introducing individual symmetry
transformations for each walker which transform the respective
Fokker-Planck operators to the \emph{same\/} Hermitian form. In the
transformed frame the problem is solved by the standard procedure
constructing the joint probability density from the antisymmetrized
product of the single-walker probability densities. This solution
enables us to effectively reduce the numerical efforts needed for
the evaluation of the joint probability density, mean coalescence
time, spatial probability density of coalescence position, etc.\ by
one dimension. Moreover, some of the quantities of interest can be
obtained this way fully analytically including all the
characteristics of the high-barrier case which are very hard to
reliably determine numerically.

The paper is organized as follows. After introducing the discrete
model in Sec.~\ref{SecRates} and its continuum limit in
Sec.~\ref{SecMEFP}, we solve the latter in Sec.~\ref{SecSolutionFP}
by using the symmetry of the problem. The main quantities of
interest, namely, the coalescence time and its distribution, as well
as the distribution of the coalescence position are obtained in
Sec.~\ref{SecBubbleCoa}. In the following Sec.~\ref{SecME}, we
introduce a direct solution of the discrete problem via the complete
master equation and a stochastic simulations scheme (Gillespie) and
compare these results to those of the continuum approximation in
Sec.~\ref{SecComparison}. In Sec.~\ref{SecBio} we address the
connection between our models and results and real biologically
relevant data. In the last Sec.~\ref{SecConclusions} we state our
conclusions. Appendix \ref{AppCalgt} presents a detailed calculation
of the auxiliary single-walker density. In Appendix
\ref{AppMEnumerical} we explain the direct numerical solution to the
full master equation of Sec.~\ref{SecME}, while in Appendix
\ref{AppGillespie} the Gillespie stochastic simulation scheme for
the same master equation is briefly summarized.

\section{Defining the zipping and unzipping rates}\label{SecRates}
In this section we define the transition rates for opening or
closing a base pair, which are determined by two effects: the
energy landscape stemming from the thermodynamical partition
factors and the thermal fluctuations.

As illustrated in Fig.~\ref{fig:model}, we consider the case when
the two soft bubble zones of the DNA construct are preferentially
open, while the central barrier region is initially completely
closed. We assume that the two soft zones are homopolymers with
identical melting temperature $T_s$, and that the barrier region is
a homopolymer with melting temperature $T_b>T_s$. In the following,
we neglect secondary structure formation in the bubbles, consistent
with experimental observations in relatively short bubble domains
\cite{altan}. The barrier region of initially closed base-pairs
between the zipper forks will also be referred to as the clamp.

Each of the two DNA-bubbles is characterized by a partition factor
of the form
\begin{eqnarray}
\label{one}
\mathscr{Z}_L(X)&=&\frac{\sigma_0}{(X+N_L+1)^{c}}
\prod_{\tilde{X}=-N_L}^{X}
u(\tilde{X}), \label{EqZLX}\\
\label{two} \mathscr{Z}_R(Y)&=&\frac{\sigma_0}{(N+N_R-Y+1)^{c}}
\prod_{\tilde{Y}=Y}^{N+N_R} u(\tilde{Y}),
\end{eqnarray}
where $X,Y\in[-N_L,N+N_R]$ denote the positions of the left and
right zipper fork, respectively. In Eqs.~(\ref{one}) and (\ref{two})
the quantity $u(X)$, which takes the values $u_{s(b)}$, is the
Boltzmann factor for breaking a base-pair in the soft zone (barrier
domain), $u_{s(b)}=\exp\left(\beta \Delta G_{s(b)}\right)$,
corresponding to the free energy $\Delta G_{s(b)}$ for breaking a
base-pair; moreover, $\beta\equiv1/[k_BT]$. We define
$u(-N_L)=u(N+N_R)\equiv1$.

Note once more the prefactor $(bubble\ length+1)^{-c}$ displaying
the inherent long-range character of the Poland-Scheraga free energy
model. It measures the reduction of the degrees of freedom of a loop
configuration, as characterized by the critical exponent $c$
\cite{santalucia,guttmann,poland,fisher1,fixman}. For the long-time behavior in
larger, single bubbles the influence of $c$ on the distribution of
bubble lifetimes is considered in Refs.~\cite{hans,kafri} in a
continuum approach.

Finally, $\sigma_0$ is the cooperativity factor corresponding to the
free energy barrier for breaking the first base-pair in a stretch of
intact double-strand. Loosely speaking, it corresponds to the
disruption of two stacking interactions in the DNA, while the single
open base-pair's entropy gain cannot balance the required enthalpy.
This contrasts the opening of further base-pairs, for which the
entropy gain almost balances the enthalpy cost. The cooperativity
factor $\sigma_0$ helps stabilizing the coalesced DNA stretch
against reclosure, as the combined free energy of the two individual
bubbles carries a factor $\sigma_0^2$ while the coalesced bubble has
only a factor $\sigma_0$.

The full partition function is
\begin{equation}\label{EqFullPartition}
\mathscr{Z}(X,Y)=\mathscr{Z}_L(X)\mathscr{Z}_R(Y).
\end{equation}
It defines the free energy landscape
$\mathscr{F}(X,Y)=-\beta^{-1}\log\left[\mathscr{Z}(X,Y)\right]$, in
which the random motion of the zipper forks takes place, as the
gradient of $\mathscr{F}$ with respect to the coordinates $X$ and
$Y$ defines the local driving forces experienced by the two
zipper forks.\\

Below the melting temperature of the barrier $T_b$, the barrier will
on average be driven towards closure, while above $T_b$ it will tend
to denature completely. The effect of thermal fluctuations is to
introduce a random walk-type dynamics of the position of the two
zipper forks. Eventually, full denaturation of the clamp may be
reached even below the melting temperature $T_b$. Once the two
bubbles coalesce, the loop initiation (cooperativity) factor
$\sigma_0$ is released, and the coalesced state becomes stabilized
against closure.

Dynamically we quantify the random motion of the two zipper forks
due to thermal fluctuations as follows. To zip close an already
opened base-pair, we assume that this process is mainly governed by
diffusion-limited encounter of the two separated bases, and
subsequent bond formation. In contrast, to unzip a still closed
base-pair, the free energy barrier embodied in the Boltzmann factor
$u$ has to be overcome. For the left zipper fork we define
$t^+_L(X,Y)$ which is the transfer coefficient for the process
$X\rightarrow X+1$, corresponding to clamp size decrease, and
$t^-_L(X,Y)$ the transfer coefficient for the process $X\rightarrow
X-1$ (clamp size increase). For the right zipper fork we similarly
introduce $t^+_R(X,Y)$ for the process $Y\rightarrow Y+1$ (clamp
size increase) and $t^-_R(X,Y)$ for the process $Y\rightarrow Y-1$
(clamp size decrease). Due to the end clamping we require that $X\ge
-N_L$ and $Y\le N+N_R$, which amounts to introducing reflecting
boundary conditions \footnote{Also, $t_L^+(X=-N_L-1,Y)=0$ and
$t_R^-(X,Y=N+N_R+1)=0$ for completeness.}
  \begin{equation}
t_L^-(X=-N_L,Y)=0,\label{Eqreflecting1}
  \end{equation}
and
  \begin{equation}
t_R^+(X,Y=N+N_R)=0.\label{Eqreflecting2}
  \end{equation}
Once the clamp has vanished, we assume that the clamp will not be
able to reform for a long time, and we impose the absorbing
conditions
  \begin{equation}
t_L^-(X,X)=t_R^+(Y,Y)=0.\label{eq:absorbing}
  \end{equation}
Dynamically, this is connected to the time it takes the long stretch
of single-strand to re-establish a base-pair in the clamp region
(diffusion limit). In terms of the free energy the suppression of
clamp reformation is due to the release of the free energy $\Delta
G_{\rm bubble}$ corresponding to the cooperativity factor $\sigma_0$
on bubble coalescence (it would cost the additional factor
$\sigma_0$ to reintroduce two single-strand/double-strand
boundaries).

Knowledge of the transfer coefficients together with the boundary
conditions above completely determine the dynamics, and we proceed
by giving explicit expressions for the transfer coefficients in
terms of the physical parameters of the problem.
%We denote by $u(x)$ the statistical weight associated with bond
%breaking of a base-pair at position $x$, see Fig.~\ref{fig:model}.
For the zipping rates we choose
  \begin{equation}
t^-_L(X,Y)=\frac{1}{2} \mathcal{K}(X+N_L),\label{eq:t_L_plus}
  \end{equation}
for the left fork, and identically for the right fork
  \begin{equation}
t^+_R(X,Y)=\frac{1}{2} \mathcal{K}(N_R+N-Y). \label{eq:t_R_minus}
  \end{equation}
We defined above a bubble-size-dependent rate coefficient
  \begin{equation}
\mathcal{K}(q)= k q^{-\mu},\label{eq:rate_coeff}
  \end{equation}
with $q$ being the number of broken bps in the bubble, where we
have, as in previous studies, introduced the hook exponent $\mu$,
related to the fact that during the zipping process not only the
base-pair at the zipper fork is moved, but also part of the
single-strand is dragged or pushed along. This additional effect may
be included using similar arguments as in
Refs.~\cite{Di_Marzio_Guttman_Hoffman,tobiasrc,tobiasjpc}: To zip
close a base-pair, the two single-strands making up the bubble have
to be pulled closer towards the zipper fork.  The adjustment of
pulling propagates along the contour of the chain until the closest
bend (inflexion) is reached, a distance that scales as the gyration
radius, i.e. $\simeq q^{\nu}$.  Having in mind Rouse-type dynamics,
this would slow down the unzipping rates by the factor $q^{-\nu}$.
Hydrodynamic interactions may change the exponent and we here take
the transfer coefficients above proportional to $q^{-\mu}$, with
$\mu$ to be determined by more detailed microscopic investigations.
The rate constant $k$ appearing in Eq.~(\ref{eq:rate_coeff}) is the
rate constant for pure base pair unzipping without factors due to
the coupling along the chain, i.e., the hook exponent. The factor
$1/2$ introduced above in Eqs.~\eqref{eq:t_L_plus},
\eqref{eq:t_R_minus} is merely for convenience to be consistent with
the nomenclature of previous approaches \cite{tobiasrc,tobiasjpc}.
Apart from the hook effect, we thus assign a factor $k/2$ for the
zipping at each of the forks.

As the DNA construct is embedded in a thermal bath, we require the
zipping rates to fulfill the detailed balance conditions
  \begin{equation}
 t_L^+(X-1,Y)\mathscr{Z}(X-1,Y) =
t_L^-(X)\mathscr{Z}(X,Y),\label{eq:det_balance_1}
  \end{equation}
and
  \begin{equation}
t_R^-(X,Y+1)\mathscr{Z}(X,Y+1)
=t_R^+(X,Y)\mathscr{Z}(X,Y).\label{eq:det_balance_2}
 \end{equation}
These conditions guarantee the relaxation to the thermodynamic equilibrium.
For the left zipping rates this is fulfilled for
  \begin{equation}
t_L^+(X,Y)=\frac{1}{2}\mathcal{K}(q_L+1)
u(X+1)\{(q_L+1)/(q_L+2)\}^c. \label{eq:t_L_minus}
  \end{equation}
Here $q_L=X+N_L$ is the length of the left bubble, and again
$(bubble\ length+1)^{-c}$ is the correction for the entropy loss of
a closed polymer loop, with $c$ being the loop exponent. For bubble
size increase we thus take the transfer coefficients to be
proportional to the Arrhenius-factor $u(X+1)$, multiplied by a loop
correction factor. Note that for $q_L\rightarrow \infty$ the loop
correction factor tends to $1$, which we will exploit later. For the
right fork we similarly find
  \begin{equation}
t_R^-(X,Y)=\frac{1}{2}\mathcal{K}(q_R+1)
u(Y-1)\{(q_R+1)/(q_R+2)\}^c, \label{eq:t_R_plus}
  \end{equation}
where $q_R=N+N_R-Y$ is the length of the right bubble. We point out
that Eqs.~(\ref{eq:t_L_plus}), (\ref{eq:t_R_minus}),
(\ref{eq:t_L_minus}), and (\ref{eq:t_R_plus}) are not unique in
satisfying the detailed balance conditions,
Eqs.~(\ref{eq:det_balance_1}) and (\ref{eq:det_balance_2}). However,
different choices correspond to redefinitions of the time unit $1/k$
which is the free parameter in our master equation approach and
needs to be fixed from fit to experiment.

From the transition rates a master equation can be constructed for
the conditional probability $P(X,Y;t|X_0,Y_0)$ with $X_0,Y_0$ being
the initial positions of the zipper forks. This master equation can
be solved numerically, details of which are introduced in
Sec.~\ref{SecME}, and physical quantities such as the
mean-first-passage time density can be calculated. However, based on
four assumptions concerning the transition rates it is possible to
derive a continuous Fokker-Planck equation approximating the full
master equation description; this is done in Sec.~\ref{SecMEFP}.
From the Fokker-Planck approach we then derive numerical results for
the coalescence time density, and both numerical and analytic
expressions for the mean coalescence time and the probability
density for the coalescence position in Secs.~\ref{SecSolutionFP}
and \ref{SecBubbleCoa}. These three following sections provide
details and extensions of our previous short work \cite{twobubb}.

\section{The Fokker-Planck
approximation to the master equation}\label{SecMEFP}

In this section we derive a Fokker-Planck approximation to the
master equation based on the following assumptions:
\begin{itemize}
\item[({\it i})] The temperature $T$ is so high compared to $T_s$ that the base
pairs in the soft zones remain unzipped at all times, i.e. there are
effectively reflecting boundary conditions at the interfaces between
the soft zones and the barrier region;
\item[({\it ii})] the soft zones are sufficiently long such that the
influence of the loop factors can be neglected, and similarly
\item[({\it iii})] the influence of the hook factors becomes sufficiently
small.
\item[({\it iv})] Finally, the number of bps in the barrier region is much bigger than one, i.e.\ $N\gg 1$, which
allows for taking the continuum limit (see below).
\end{itemize}

\begin{figure}
\scalebox{1}{\input{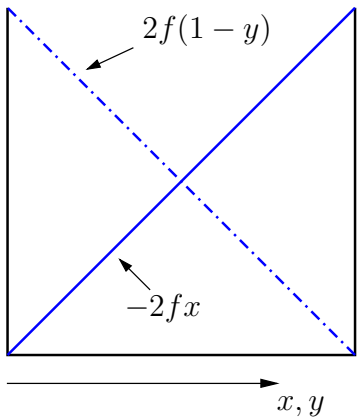}} \caption{Plot of the
linear potentials experienced by the respective bubble interfaces in
the case $T<T_b\ (f<0)$ in terms of the dimensionless quantities
$x,y,f$ (see text for details). } \label{fig:modelFP}
\end{figure}

Under the assumptions ({\it i})--({\it iii}) the full partition
function for the two bubbles and the partially denatured barrier
region becomes [see Eq.~(\ref{EqFullPartition})]
\begin{equation}
\label{part} \mathscr{Z}=\sigma_0^2u_s^{N_L+N_R}u_b^{X+N-Y},
\end{equation}
where in this section $X,Y\in\{0,1,2,\dots,N\}$ due to ({\it i})
with $X\leq Y$. Notice that $X$ is the number of barrier base-pairs
already broken from the left end of the barrier, and $N-Y$ counts
the broken barrier base-pairs from the right end. The free energy is
given as
\begin{equation}
\mathscr{F}=-2k_BT\log\sigma_0-(N_L+N_R)\Delta G_s+(X+N-Y)\Delta
G_b.
\end{equation}
In the continuum limit (assumption ({\it iv})) we introduce
dimensionless coordinates $x=\tfrac{X}{N}$ and $y=\tfrac{Y}{N}$,
$x,y\in[0,1]$. The gradient of $\mathscr{F}$ with respect to the
coordinates $x$ and $y$ defines the local force experienced by the
two zipper forks, namely
\begin{equation}
F_X=-\frac{d\mathscr{F}}{dx}=-N \Delta
G_b,\,\,F_Y=-\frac{d\mathscr{F}}{dy}= N \Delta G_b=-F_X,
\end{equation}
and we immediately see, that the zipper forks $X$ and $Y$ are driven
by opposite, constant forces as sketched in
Fig.~\ref{fig:modelFP}.

Using the simplifications ({\it ii}) and ({\it iii}) stated above,
the modified continuum rates (denoted by $r$ to distinguish them
from the notation introduced in the discussion of the discrete
case) for closing a base-pair at the left fork at position $x$, or
at the right fork at position $y$, become [see
Eqs.~(\ref{eq:t_L_plus}), (\ref{eq:t_R_minus}),
(\ref{eq:t_L_minus}), and (\ref{eq:t_R_plus})]
\begin{equation}
r_L^-(x,y)=r_R^+(x,y)=k/2,
\end{equation}
and
\begin{equation}
r_L^+(x,y)=r_R^-(x,y)=u_bk/2,
\end{equation}
such that the zipping open of a base-pair requires crossing the
barrier $\Delta G_b$. The boundary conditions
\begin{equation}
r_L^-(0,y)=r_R^+(x,1)=r_L^+(x,x)=r_R^-(y,y)=0,
\end{equation}
guarantee that base-pairs cannot close beyond the barrier region, and that
the process ends when the two zipper forks coalesce.

Define by $P(x,y;\tau)$ the probability distribution that the left
and right zipper forks are located at $x$ and $y$, respectively, at
some given time $\tau$. The time evolution of $P(x,y;\tau)$ is then
given in terms of the master equation \cite{van_Kampen,Risken}
\begin{widetext}
\begin{eqnarray}
\nonumber \frac{\partial}{\partial
\tau}P(x,y;\tau)&=&r_L^+(x-1/N,y)P(x-1/N,y;\tau)
+r_L^-(x+1/N,y)P(x+1/N,y;\tau)-\left[r_L^-(x,y)+r_L^+(x,y)\right]P(x,y;\tau)\\
&&\hspace*{-1.2cm}
+r_R^-(x,y+1/N)P(x,y+1/N;\tau)+r_R^+(x,y-1/N)P(x,y-1/N;\tau)-\left[r_R^-(x,y)+r_R^+(x,y)
\right]P(x,y;\tau). \label{master}
\end{eqnarray}
Following the standard derivation \cite{Gardiner,van_Kampen} we
Taylor-expand the above master equation keeping the first two orders
only. For instance, for the first term on the right hand side of
Eq.~(\ref{master}), we obtain the Taylor expansion
\begin{equation}
r_L^+(x-1/N,y)P(x-1/N,y;\tau)\approx
r_L^+(x,y)P(x,y;\tau)-\frac{1}{N}\frac{\partial}{
\partial x}r_L^+(x,y)P(x,y;\tau)+\frac{1}{2N^2}\frac{\partial^2}{\partial
x^2}r_L^+(x,y)P(x,y;\tau).
\end{equation}
This is the only consistent expansion of finite order according to
the Pawula-Marcinkiewicz theorem
\cite{Risken,pawula,marcienkiewicz}. Alternatively the full
Kramers-Moyal expansion needs to be taken along. With analogous
expansions for the other terms and after some rearrangement, we find
the bivariate Fokker-Planck equation \cite{Risken}
\begin{equation}
\frac{\partial}{\partial \tau}P(x,y;\tau|x_0,y_0)=
F\left(\frac{\partial}{\partial y}-\frac{
\partial}{\partial x}\right)P(x,y;\tau|x_0,y_0)+D\left(\frac{\partial^2}{\partial
x^2}+\frac{\partial^2}{\partial y^2}\right)P(x,y;\tau|x_0,y_0),
\label{fpe}
\end{equation}
\end{widetext}
where, instead of the probability density $P(x,y;\tau)$, we use
explicitly the notation $P(x,y;\tau|x_0,y_0)$ including the initial
conditions $x_0$ and $y_0$. In Eq.~(\ref{fpe}), the force $F$ and
diffusion constant $D$ are defined by
\begin{equation}
F\equiv\frac{k(u_b-1)}{2N},
\end{equation}
and
\begin{equation}
D\equiv\frac{k(u_b+1)}{4N^2}.
\end{equation}
Eq.~(\ref{fpe}) is completed by specifying the initial and boundary conditions.
As initial condition, we choose the sharp $\delta$-form
\begin{equation}\label{EqInitialCond}
P(x,y;0|x_0,y_0)=\delta(x-x_0)\delta(y-y_0),
\end{equation}
with $x_0<y_0$, and due to the initial condition
Eq.~(\ref{EqInitialCond}) the joint probability density
$P(x,y;\tau|x_0,y_0)$ is actually the Green's function of
Eq.~(\ref{fpe}). The condition that the two bubbles in the soft
zones are always open is guaranteed by the reflecting boundary
conditions (here, we define $2f\equiv F/D$)
\begin{eqnarray}
\nonumber &&\left.\left(\frac{\partial}{\partial
x}-2f\right)P(x,y;\tau|x_0,y_0)
\right|_{x=0}=0,\\
&&\left.\left(\frac{\partial}{\partial
y}+2f\right)P(x,y;\tau|x_0,y_0) \right|_{y=1}=0,
\end{eqnarray}
at the edges of the line segment $[0,1]$: Once a zipper fork reaches
either edge, the only possible direction to move is to restart
unzipping the barrier. Moreover, we specify the viciousness
condition \footnote{The name {\it vicious} stems from
Ref.~[\onlinecite{fisher}].}
\begin{equation}
\label{vicious} P(x,x;\tau|x_0,y_0)=0,
\end{equation}
according to which the two zipper forks cannot be at the same
point: the two forks annihilate and the bubbles coalesce.
This last conditions ensures the continuous character of
the probability density $P$. For completeness, we actually need to
specify a second set of boundary conditions. However, due to the
viciousness condition (\ref{vicious}), we can choose this boundary
condition \emph{ad libitum}; a clever choice will turn out to be
\begin{eqnarray}
\nonumber
&&\left.\frac{\partial}{\partial x}P(x,y;\tau|x_0,y_0)\right|_{x=1}=0,\\
&&\left.\frac{\partial}{\partial
y}P(x,y;\tau|x_0,y_0)\right|_{y=0}=0.
\end{eqnarray}
Such a choice is possible because the zipper forks never reach these
two points.

\section{Solution of the vicious walker
problem}\label{SecSolutionFP}

\subsection{Transformation of the Fokker-Planck equation}

To obtain the solution of the Fokker-Planck equation (\ref{fpe}), it
is convenient to notice that after a redefinition of time unit
$t=D\tau$ the problem depends on a single dimensionless parameter
\begin{equation}\label{def_f}
f =\frac{F}{2D}=N\frac{u_b-1}{u_b+1}.
\end{equation}
It is important that this parameter depends on the length of the
barrier $N$ and the Boltzmann factor $u_b$ for opening the barrier
bps but not on the kinetic constant $k$. Thus, apart from an overall
prefactor fixing the time unit, the solution depends solely on the
structural properties of the physical system under study.

Let us summarize the rephrased problem in terms of $f$ for completeness
\begin{subequations}
\begin{align}
\label{redfpe} \left[-\frac{\partial}{\partial
t}+\frac{\partial^2}{\partial x^2}+\frac{\partial^2}{\partial y^2}
-2f \frac{\partial}{\partial
x}+2f\frac{\partial}{\partial y} \right]&P(x,y;t|x_0,y_0)=0,\\
\intertext{with boundary conditions} \label{reflectx}
\left.\left(\frac{\partial}{\partial x}-2f\right)P(x,y;t|x_0,y_0)
\right|_{x=0}&=0,\\
\left.\left(\frac{\partial}{\partial y}+2f\right)P(x,y;t|x_0,y_0)
\right|_{y=1}&=0, \label{reflecty}\\
\label{completex}\left.\frac{\partial}{\partial x}P(x,y;t|x_0,y_0)\right|_{x=1}&=0,\\
\label{completey}\left.\frac{\partial}{\partial
y}P(x,y;t|x_0,y_0)\right|_{y=0}&=0,\\
\intertext{the viciousness condition}
\label{EqVicious} P(x,x;t|x_0,y_0)= 0,&\\
\intertext{and the ininitial condition} \label{init_cond}
P(x,y;0|x_0,y_0)=\delta(x-x_0)\delta(y-y_0)& \text{ with } x_0<y_0.
\end{align}
\end{subequations}

To proceed in the solution, let us first introduce the Fokker-Planck
operators
\begin{subequations}
\begin{eqnarray}
\mathbb{L}_{\mathrm{FP}}^+(x)\equiv\frac{\partial^2}{\partial x^2}-2f
\frac{\partial}{\partial x},\\
\mathbb{L}_{\mathrm{FP}}^-(y)\equiv\frac{\partial^2}{\partial y^2}+2f
\frac{\partial}{\partial y},
\end{eqnarray}
\end{subequations}
so that the Fokker-Planck equation \eqref{redfpe} can be recast into
the following form
\begin{equation}
\frac{\partial}{\partial
t}P(x,y;t|x_0,y_0)=\Big[\mathbb{L}_{\mathrm{FP}}^+(x)
+\mathbb{L}_{\mathrm{FP}}^-(y)\Big]P(x,y;t|x_0,y_0).
\end{equation}
The operators $\mathbb{L}_{\mathrm{FP}}^+(x)$ and
$\mathbb{L}_{\mathrm{FP}} ^-(y)$ can now be used to transform our
Fokker-Planck equation following the procedure outlined in
Ref.~\cite{Risken}, Chapter 5.4. Let us now define the Hermitian
operator
\begin{equation}
\label{EqLoperator} \mathbb{L}(x,x_0)\equiv
e^{-f(x-x_0)}\mathbb{L}_{\mathrm{FP}}^+(x)e^{f(x-x_0)}
=\frac{\partial^2}{\partial x^2}-f^2.
\end{equation}
Here the last equality sign can be shown by applying the operator to
a test function. The operator $\mathbb{L}(x,x_0)$ corresponds to a
Hamilton operator of a Schr{\"o}dinger equation with imaginary time
$-i\hbar t$, mass $m=\hbar^2/2$ and a constant potential $f^2$. The
idea is that it is (in general) easier to solve the time-dependent
Schr\"{o}dinger equation than the Fokker-Planck equation, because
the first-order derivative has been eliminated. The relation between
the solutions of the original and the transformed Fokker-Planck
equations are found after a few lines of algebra, and we obtain the
following result: If $\widetilde{p}^L(x;t|x_0)$ is a solution of
\begin{equation}
\frac{\partial}{\partial
t}\widetilde{p}^L(x;t|x_0)=\mathbb{L}(x,x_0)
\widetilde{p}^L(x;t|x_0),
\end{equation}
then the density
\begin{equation}
\label{similarity}
p^L(x;t|x_0)=\exp[f(x-x_0)]\widetilde{p}^L(x;t|x_0)
\end{equation}
solves the original equation
\begin{equation}
\frac{\partial}{\partial
t}p^L(x;t|x_0)=\mathbb{L}_{\mathrm{FP}}^+(x)p^L (x;t|x_0).
\end{equation}
The boundary conditions in $x$, Eqs.~(\ref{reflectx}) and (\ref{completex}),
are transformed to
\begin{subequations}
\begin{eqnarray}
&&\left.\left(\frac{\partial}{\partial
x}-f\right)\widetilde{p}^L(x;t|x_0)
\right|_{x=0}=0, \label{EqBoundxLgt}\\
&&\left.\left(\frac{\partial}{\partial
x}+f\right)\widetilde{p}^L(x;t|x_0) \right|_{x=1}=0,
\label{EqBoundxRgt}
\end{eqnarray}
\end{subequations}
while the initial condition remains unchanged,
$\widetilde{p}^L(x;t=0|x_0)=
\exp[-f(x-x_0)]p^L(x;t=0|x_0)=\delta(x-x_0)$. In other words, with
Eqs.~\eqref{EqLoperator}, \eqref{EqBoundxLgt}, and
\eqref{EqBoundxRgt} we have indeed transformed the original
Fokker-Planck operator $\mathbb{L}_{\mathrm{FP}}^+(x)$ into
Hermitian form.

Noticing that
\begin{equation}
\label{EqLyoperator} \mathbb{L}(y,y_0) = \frac{\partial^2}{\partial
y^2}-f^2 = e^{f(y-y_0)}\mathbb{L}_{\mathrm{FP}}^-(y)e^{-f(y-y_0)},
\end{equation}
the same procedure can be carried out for the $y$-coordinate, where
the different sign in front of $f$ in Eq.~(\ref{redfpe}) causes
that, if $\widetilde{p}^R(y;t|y_0)$ is a solution to
\begin{equation}
\frac{\partial}{\partial
t}\widetilde{p}^R(y;t|y_0)=\mathbb{L}(y,y_0)
\widetilde{p}^R(y;t|y_0),
\end{equation}
then
\begin{equation}\label{similarity1}
p^R(y;t|y_0)=\exp\left[-f(y-y_0)\right]\widetilde{p}^R(y;t|y_0)
\end{equation}
satisfies
\begin{equation}
\frac{\partial}{\partial
t}p^R(y;t|y_0)=\mathbb{L}_{\mathrm{FP}}^-(y) p^R(y;t|y_0),
\end{equation}
and the boundary conditions are
\begin{subequations}
\begin{eqnarray}
&&\left.\left(\frac{\partial}{\partial
y}+f\right)\widetilde{p}^R(y;t|y_0)
\right|_{y=1}=0, \label{EqBoundyRgt} \\
&&\left.\left(\frac{\partial}{\partial
y}-f\right)\widetilde{p}^R(y;t|y_0) \right|_{y=0}=0,
\label{EqBoundyLgt}
\end{eqnarray}
\end{subequations}
together with $\tilde{p}^R(y;t=0|y_0)=\delta(y-y_0)$.

We now see why Eqs.~\eqref{completex}, \eqref{completey} are clever
choices for the additional boundary conditions on
$P(x,y;t|x_0,y_0)$, together with the similarity transformations
\eqref{similarity} and \eqref{similarity1}: reflecting the uneven
symmetry of the problem with respect to the original coordinates,
\emph{all\/} equations defining the functions
$\widetilde{p}^{L}(x;t|x_0)$ and $\widetilde{p}^{R} (x;t|x_0)$ are
identical, and thus are the functions themselves
\begin{equation}
\tilde{p}^{L}(x;t|x_0)=\tilde{p}^{R}(x;t|x_0)=\tilde{p}(x;t|x_0).
\end{equation}\\

\subsection{Solution of the transformed Fokker-Planck equation}

After the individual similarity transformations for the left
\eqref{similarity} and right \eqref{similarity1} walker,
respectively, are performed, we arrive at the following
time-dependent Schr{\"o}dinger equation
\begin{equation}
\label{EqSchrodOrg} \frac{\partial}{\partial
t}\widetilde{P}(x,y;t|x_0,y_0)=\left[\mathbb{L}
(x,x_0)+\mathbb{L}(y,y_0)\right]\widetilde{P}(x,y;t|x_0,y_0),
\end{equation}
with imaginary time; here,
\begin{equation}\label{similarity2}
\widetilde{P}(x,y;t|x_0,y_0)=e^{-f(x-x_0)+f(y-y_0)}P(x,y;t|x_0,y_0),
\end{equation}
which implies the same viciousness condition
$\widetilde{P}(x,x;t|x_0,y_0)=0$ as in the original formulation.
Now, however, we effectively have two identical vicious walkers
moving in a common (constant) potential for which the solution is
well known \cite{fisher,bray}. It is given by the antisymmetric
product of the single-walker solutions, namely
\begin{equation}\label{EqSchrodSol}
    \widetilde{P}(x,y;t|x_0,y_0)=\tilde{p}(x;t|x_0)\tilde{p}(y;t|y_0)-\tilde{p}(y;t|x_0)\tilde{p}(x;t|y_0).
\end{equation}
Note that this form of the solution is analogous to constructing the
solution of an absorbing boundary value problem for a single
diffusor under a constant drift according to the method of images
\cite{redner}.

The backward transformation of the solution (\ref{EqSchrodSol}) by
inverting Eq.~\eqref{similarity2} finally produces
\begin{widetext}
\begin{equation}\label{green}
P(x,y;t|x_0,y_0)=e^{f(x-x_0)-f(y-y_0)}\left[\tilde{p}(x;t|x_0)\tilde{p}(y;t|y_0)
-\tilde{p}(y;t|x_0)\tilde{p}(x;t|y_0)\right],
\end{equation}
and by construction this is a solution of Eq.~(\ref{redfpe}),
satisfying the boundary conditions
Eqs.~(\ref{reflectx})--(\ref{completey}), as well as the viciousness
condition $P(x,x;t|x_0,y_0)=0$. It remains to check that this
solution also satisfies the initial condition \eqref{init_cond} and,
thus,  fully solves the studied problem. For $t=0$ we obtain
\begin{eqnarray}
\nonumber
P(x,y;t=0|x_0,y_0)&=&e^{f(x-x_0)-f(y-y_0)}\left[\delta(x-x_0)\delta(y-y_0)
-\delta(y-x_0)\delta(x-y_0)\right]\\
\nonumber
&=&\delta(x-x_0)\delta(y-y_0)-e^{2f(y_0-x_0)}\delta(y-x_0)\delta(x-y_0)\\
&=&\delta(x-x_0)\delta(y-y_0),
\end{eqnarray}
\end{widetext}
valid for  $x_0<y_0$ and $x<y$. The last equality follows because of
the choice of the initial condition $x_0<y_0$ and the viciousness of
the process, i.e.\ the walkers can never pass each other, which also
implies $x<y$ for any realizable configuration at any time $t$.
Thus, the arguments of the $\delta$-functions in the second term can
never vanish simultaneously and therefore this term is effectively
equal to zero. Thus \eqref{green} is the solution to our problem.

\subsection{Calculation of $\tilde{p}(x;t|x_0)$ and its spectral resolution}
\label{SecCalgt}

To find the solution of the full 2-walker problem \eqref{green} we
need to calculate $\widetilde{p}(x;t|x_0)$, which solves the
Schr{\"o}dinger equation
\begin{equation}\label{seq}
\frac{\partial}{\partial t} \widetilde{p}(x;t|x_0)=
\left[\frac{\partial^2}{\partial
x^2}-f^2\right]\widetilde{p}(x;t|x_0),
\end{equation}
and satisfies the boundary conditions Eqs.~(\ref{EqBoundxLgt}) and
(\ref{EqBoundxRgt}), as well as the initial condition
$\widetilde{p}(x;t=0|x_0)=\delta(x-x_0)$. Unfortunately, there is no
explicit solution of this equation in the time domain. It can be
found, however, in the Laplace picture which is done in detail in
Appendix \ref{AppCalgt}. Here we only summarize the final result
\begin{eqnarray}
\nonumber \tilde{p}(x;z|x_0)&=&\frac{1}{2k(\kappa^2
e^{2k}-1)}\left[e^{k|x-x_0|}
+\kappa^2 e^{2k} e^{-k|x-x_0|}\right.\\
&&\left.+\kappa e^{(x+x_0)k}+\kappa e^{2k} e^{-(x+x_0)k}\right],
\label{1partGF}
\end{eqnarray}
with $z$-dependent
\begin{equation}
k\equiv k(z)=\sqrt{z+f^2},
\end{equation}
and
\begin{equation}
\kappa\equiv\kappa(z)=\frac{k(z)+f}{k(z)-f}.
\end{equation}
The corresponding behavior in the
time domain is found by the inverse Laplace transform
\begin{eqnarray}
\nonumber
\tilde{p}(x;t|x_0)&=&\int_{\epsilon-i\infty}^{\epsilon+i\infty}\frac{dz}{2\pi
i}e^{zt} \tilde{p}(x;z|x_0)\\
&=&\sum_{n=0}^{\infty}e^{\lambda_n t}\psi_n(x)\psi^*_n(x_0)
\label{eigen_expansion}
\end{eqnarray}
with $\epsilon\ge 0$ large enough such that all singularities of
$\tilde{p}(x;z|x_0)$ lie in the half-plane
$\mathrm{Re}\,z<\epsilon$. We used the formal eigenmode expansion in
the second line of the above equation which exists as a spectral
resolution of the Hermitian operator defined by Eq.~(\ref{seq}) and
the pertinent boundary conditions (\ref{EqBoundxLgt}) and
(\ref{EqBoundxRgt}). The eigenvalues $\lambda_n$ are real numbers
since the operator is Hermitian, but not necessarily non-positive
like in the case of standard Fokker-Planck operators in one
dimension. The reason is that $\tilde{p}(x;t|x_0)$ is an auxiliary
mathematical quantity without any direct physical meaning and, thus,
can in principle grow exponentially over time, i.e., some of the
$\lambda_n$ might be positive. The physical quantity which is not
allowed to grow indefinitely is the 2-walker Green's function
(\ref{green}). This is useful to keep in mind when studying the
spectral resolutions of \eqref{1partGF} and \eqref{green} in more
detail in the following.

The spectrum of the operator $\partial^2/\partial x^2-f^2$ from
Eq.~(\ref{seq}), together with the boundary conditions, can be found
either directly from the defining equations or by finding the poles
of the 1-walker Green's function \eqref{1partGF}. Indeed, the
secular equation obtained by either method is equivalent to the
denominator in Eq.~\eqref{1partGF} being zero, i.e.,
$\kappa^2(\lambda)= e^{-2k(\lambda)}$. This leads to transcendent
equations for real $\lambda$ in the respective ranges
\begin{subequations}
\begin{widetext}
\label{seculareq}
\begin{eqnarray}
\label{seculareq1}
(\lambda+2f^2)\sinh\sqrt{\lambda+f^2}+2f\sqrt{\lambda+f^2}\cosh\sqrt{\lambda
+f^2}&=0\quad\text{for }\lambda\ge -f^2\\
(\lambda+2f^2)\sin\sqrt{|\lambda|-f^2}+2f\sqrt{|\lambda|-f^2}\cos\sqrt{|
\lambda|-f^2}&=0\quad\text{for }\lambda\le -f^2\ .
\label{seculareq2}
\end{eqnarray}
\end{widetext}
\end{subequations}

One looks for solutions of these equations in their ranges of
validity. Thus, with the help of the standard graphical analysis, it
turns out that the first equation \eqref{seculareq1} has at least
one solution only for $f<0$, this solution is positive, i.e.\
$\lambda_0>0$. When $f<-2$ a further solution with
$-f^2<\lambda_1<0$ solves Eq.~\eqref{seculareq1}. There are no more
options for Eq.~\eqref{seculareq1}. On the other hand, the second
equation \eqref{seculareq2} always has an infinite number of
solutions. For $f>0$ these solutions are bounded by
$-f^2-(n+1)^2\pi^2\leq \lambda_n\leq -f^2-n^2\pi^2$
($n=0,1,2,\dots$). For $0>f>-2$ the first eigenvalue $\lambda_0$
satisfies Eq.~\eqref{seculareq1} while the remaining eigenvalues
($\lambda_n$ for $n=1,2,\dots$) stem from Eq.~\eqref{seculareq2} and
are still bounded by $-f^2-n^2\pi^2\leq \lambda_n\leq
-f^2-(n-1)^2\pi^2$. Finally, for $f<-2$ the first two eigenvalues
$\lambda_0$ and $\lambda_1$ are determined by Eq.~\eqref{seculareq1}
and the rest stems from Eq.~\eqref{seculareq2} being bounded by
$-f^2-n^2\pi^2\leq \lambda_n\leq -f^2-(n-1)^2\pi^2$ for
$n=2,3,\dots$.

There are two special values of the force $f=0,-2$ where the
spectrum appears to change its analytic structure. First, the $f=0$
case corresponds to the problem of two vicious walkers freely
diffusing in an impenetrable well, i.e.~in a common potential. For
this case, we do not need to use our trick as it is solved already
by a previous study \cite{bray}: The single-walker spectrum reads
$\lambda_n=(n\pi)^2$ for $n=0,1,2,\dots$ The second case, $f=-2$,
does not appear to be in any way particular physically. Actually,
neither of the two cases have any exceptional physical properties
--- all the physical quantities change smoothly across these two
points when changing $f$. The apparent singularities occur only in
the auxiliary quantities. Technically, these two cases are the only
ones where the trivial solution $\lambda=-f^2$ of
Eqs.~\eqref{seculareq} corresponds to a non-trivial, i.e.~non-zero,
solution for the eigenfunction. Mathematically, this is reflected by
the formal failure when $k(z)=0$ of the method used in Appendix
\ref{AppCalgt} leading to Eq.~\eqref{1partGF}. In such cases, the
two fundamental solutions $e^{\pm k x}$ of Eq.~\eqref{gf} become
identical and equal to a constant. The second independent solution
is then linear in $x$ according to the elementary theory of linear
differential equations. Thus, the two special cases are not directly
covered by the general solution presented in App.~\ref{AppCalgt} and
all formulas stemming from it. We do not give explicit solutions for
those two singular cases since all quantities of interest can be
obtained from the general formulas by taking the appropriate limit
$f\to 0$, or $-2$.

Now, if we insert the expansion \eqref{eigen_expansion} into
Eq.~(\ref{green}) we see that the spectrum of that equation is given
by all pairwise sums of \emph{different\/} 1-walker eigenvalues,
i.e.\ $\Lambda_k=\lambda_i+\lambda_j\ (i\neq j)$ [the sums of
identical 1-walker eigenvalues have zero weight due to the
anti-symmetrization in (\ref{green})]. Therefore, the first two
1-walker solutions $\lambda_0$ and $\lambda_1$ determine the
long-time asymptotics of the 2-walker problem. For large negative
$f\ll -1$, corresponding to a large barrier case, the asymptotic
behavior is dominating the whole solution and, thus, the combination
$\Lambda_0\equiv\lambda_0+\lambda_1$ effectively determines the
interesting quantities such as the mean coalescence time or the
probability distribution of the coalescence position. This allows us
to get full analytic results in the limit $f\ll-1$, as explicitly
derived in Section~\ref{SecResults}. In this case both eigenvalues
$\lambda_{0,1}$ are solutions of Eq.~(\ref{seculareq1}), and they
are exponentially small in $|f|$. They can be found to leading order
from the second order expansion in $\lambda$ of
Eq.~\eqref{seculareq1}, yielding $\lambda_{0,1}\simeq\pm
4f^2e^{-|f|}$. To obtain $\Lambda_0$ we need to increase the
accuracy by expanding the equation up to the third order in
$\lambda$ and we find $\Lambda_0=\lambda_0+\lambda_1\simeq
-16f^2(|f|-1)e^{-2|f|}$ which is negative, as it should be since
$P(x,y;t|x_0,y_0)$ cannot grow exponentially, $P$ being a physical
quantity. So, despite the existence of a positive 1-walker
eigenvalue $\lambda_0$, the physically relevant combination
$\lambda_0+\lambda_1<0$ ensures meaningful results.
\newline

\section{Bubble coalescence time and position from the solution of the Fokker-Planck equation}\label{SecBubbleCoa}

\subsection{General treatment}
\label{general}

From the solution of the Fokker-Planck equation, we now calculate in
general the quantities of interest, namely the characteristic
coalescence time of the two random walking zipper forks, and the
probability distribution of the coalescence position. In the time
domain this is done by considering the conservation of probability
in the form
\begin{equation}
    \Pi(t|x_0,y_0)+\int_0^1 dy\int_0^{y}dx P(x,y;t|x_0,y_0)=1,
\end{equation}
which is actually a defining equation for the probability
$\Pi(t|x_0,y_0)$ that the walkers have met before time $t$.
Consequently, the second term represents the probability of having
two separate bubbles at time $t$ (survival probability). Note the
range of integration restricting $x\in[0,y)$. The probability
density associated with $\Pi(t|x_0,y_0)$ is
\begin{equation}
\begin{split}
\pi(t|x_0,y_0)&\equiv\frac{d}{dt}\Pi(t|x_0,y_0)\\
&=-\int_0^1 dy\int_0^y dx\frac{\partial}{
\partial t}P(x,y;t|x_0,y_0),
\end{split}
\end{equation}
which, after invoking Eq.~(\ref{redfpe}) and rearranging the order
of the integrals, yields
\begin{widetext}
\begin{equation}
\begin{split}
\pi(t|x_0,y_0)&=-\int_0^1 dy\int_0^y dx
\left[\frac{\partial^2}{\partial x^2}-2f\frac{\partial}{\partial
x}\right]P(x,y;t|x_0,y_0)-\int_0^1 dx\int_x^1 dy
\left[\frac{\partial^2}{\partial y^2}+2f \frac{\partial}{\partial
y}\right]P(x,y;t|x_0,y_0)\\
&=-\int_0^1 dy \left[\left\{\frac{\partial}{\partial
x}-2f\right\}P(x,y;t|x_0,y_0)\right]_0^y-\int_0^1 dx
\left[\left\{\frac{\partial}{\partial y}+2f
\right\}P(x,y;t|x_0,y_0)\right]_x^1\\
&=-\int_0^1 dy \left\{\frac{\partial}{\partial
x}-2f\right\}P(x,y;t|x_0,y_0) \Bigg|_{x=y}+\int_0^1 dx
\left\{\frac{\partial}{\partial y}+2f\right\}P(x,y;t|x_0,y_0)\Bigg|_{y=x}\\
&=\int_0^1 dx \left\{\frac{\partial}{\partial
y}-\frac{\partial}{\partial x} + 4
f\right\}P(x,y;t|x_0,y_0)\Bigg|_{y=x} =\int_0^1 dx
\left\{\frac{\partial}{\partial y}-\frac{\partial}{\partial
x}\right\}P(x,y;t|x_0,y_0)\Bigg|_{y=x}, \label{Eqpi}
\end{split}
\end{equation}
\end{widetext}
where the boundary conditions imposed on $P(x,y;t|x_0,y_0)$ and the
viciousness condition have been used in subsequent manipulations.
The quantity $\pi(t|x_0,y_0)$ can be interpreted as the probability
current flowing into the absorbing boundary, here given by the line
$x=y$ (compare with an analogous discussion for a 1-dimensional case
in Ref.~\cite{Novotny2000}).

From the last relation, we define the quantity
$\varrho(x;t|x_0,y_0)$, the coalescence-position-resolved
probability density for the coalescence time,
\begin{equation}
\pi(t|x_0,y_0)=\int_0^1
dx\,\varrho(x;t|x_0,y_0),\label{boundary_def1}
\end{equation}
such that \\ \\
\begin{equation}
\varrho(x;t|x_0,y_0) =\left\{\frac{\partial}{\partial
y}-\frac{\partial}{\partial
x}\right\}P(x,y;t|x_0,y_0)\Bigg|_{y=x}\label{boundary_def2}.
\end{equation}
We are mainly interested in either the mean coalescence time
$\tau(x_0,y_0)=\int_0^{\infty}dt\, t\, \pi(t|x_0,y_0)$ or the
probability density of the coalescence position
$\rho(x|x_0,y_0)=\int_0^{\infty}dt\,\varrho(x;t|x_0,y_0)$. These are
quantities integrated over time and, thus, they may be determined
from the solution in the Laplace domain without explicit knowledge
of $P(x,y;t|x_0,y_0)$ in the time domain.

We now rewrite Eq.~\eqref{green} in terms of the inverse Laplace
transforms as follows
\begin{widetext}
\begin{equation}\label{invLaplace}
P(x,y;t|x_0,y_0)=\int_{\epsilon-i\infty}^{\epsilon+i\infty}\frac{dz_1}{2\pi
i}e^{z_1t}\int_{\epsilon-i\infty}^{\epsilon+i\infty}\frac{dz_2}{2\pi
i}e^{z_2t}
e^{f(x-y-x_0+y_0)}[\tilde{p}(x;z_1|x_0)\tilde{p}(y;z_2|y_0)-
\tilde{p}(y;z_1|x_0)\tilde{p}(x;z_2|y_0)],
\end{equation}
where $\tilde{p}(x;z|x_0)$ is given in \eqref{1partGF}. In case that
$\tilde{p}(x;z|x_0)$ has no poles in the half-plane ${\rm Re}z>0$
(corresponding to $f>0,\, \epsilon=0$), we can directly use the
substitution $z_1=z/2+i\omega$, $z_2=z/2-i\omega$; $z=z_1+z_2$, and
$\omega=(z_1-z_2)/(2i)$, to obtain a single inverse Laplace
transform for $P(x,y;t|x_0,y_0)$, namely
\begin{eqnarray}
\nonumber P(x,y;t|x_0,y_0)=\int_{-i\infty}^{i\infty}\frac{dz}{2\pi
i}e^{zt}\int_
{-\infty}^{\infty}\frac{d\omega}{2\pi}e^{f(x-y-x_0+y_0)}&&\left[\tilde{p}
\left(x;z/2+i\omega|x_0\right)\tilde{p}\left(y;z/2-
i\omega|y_0\right)\right.\\
&&\left.-\tilde{p}\left(y;z/2+i\omega|x_0\right)\tilde{p}\left(
x;z/2-i\omega|y_0\right)\right].
\end{eqnarray}
Thus, we can identify the Laplace transform
$\tilde{P}(x,y;z|x_0,y_0)$ of $P(x,y;t|x_0, y_0)$ for $f>0$ as
\begin{equation}
\label{invLapf>0}
\tilde{P}(x,y;z|x_0,y_0)=\int_{-\infty}^{\infty}\frac{d\omega}{2\pi}e^{f(x-y-x_0+
y_0)}[\tilde{p}(x;z/2+i\omega|x_0)\tilde{p}(y;z/2-i\omega|y_0)-\tilde{p}(
y;z/2+i\omega|x_0)\tilde{p}(x;z/2-i\omega|y_0)].
\end{equation}
\end{widetext}
This directly yields, by means of relations (\ref{boundary_def1})
and \eqref{boundary_def2}, the Laplace transforms
$\tilde{\pi}(z|x_0,y_0)$ and $\tilde{\varrho}(x;z|x_0,y_0)$, from
which we in turn deduce the (time-averaged) distribution of
coalescence positions,
\begin{equation}\label{abco}
\rho(x|x_0,y_0)=\tilde{\varrho}(x;z=0^+|x_0,y_0),
\end{equation}
and the characteristic (mean) coalescence time
\begin{equation}\label{mfpt}
\tau(x_0,y_0)=-\left.\frac{d}{dz}\tilde{\pi}(z|x_0,y_0)\right|_{z=0^+}.
\end{equation}

For $f<0$ the situation is more complicated since
$\tilde{p}(x;z|x_0)$ in this case has a pole at $\lambda_0>0$ which
prohibits us from just repeating the above reasoning. The Laplace
transform \eqref{invLaplace} only holds for ${\rm
Re}\,z_{1,2}>\lambda_0>0$ implying ${\rm Re}\,z>2\lambda_0>0$ and
the analytic continuation down to $z=0$ is not obvious. However,
since we know the position of the only pole located in the ${\rm
Re}\,z>0$ half-plane, we can treat this singularity separately and
thus generalize the previous results. Using the Cauchy theorem for
complex integrals we move the integration line from
$\epsilon+i\omega$ with $\epsilon>\lambda_0>0$ down to the imaginary
axis and add the contribution from the (single) singularity at
$\lambda_0$ in between these two lines. This method eventually leads
to an expression for the Laplace transform $P(x,y;z|x_0,y_0)$ in the
whole half-plane $\mathrm{Re}\,z>0$ for $f<0$,
\begin{widetext}
\begin{equation}
\label{invLapf<0}
\begin{split}
    \tilde{P}(x,y;z|x_0,y_0)&=e^{f(x-y-x_0+y_0)}\bigg\{\int_{-\infty}^{\infty}\frac{d\omega}{2\pi}[\tilde{p}(x;z/2+i\omega|x_0)\tilde{p}(y;z/2-i\omega|y_0)
    -\tilde{p}(y;z/2+i\omega|x_0)\tilde{p}(x;z/2-i\omega|y_0)]\\
    &+{\rm Res}_{z_1=\lambda_0}\tilde{p}(x;z_1|x_0)\tilde{p}(y;z-\lambda_0|y_0)-{\rm
    Res}_{z_1=\lambda_0}\tilde{p}(y;z_1|x_0)\tilde{p}(x;z-\lambda_0|y_0)\\
    &+\tilde{p}(x;z-\lambda_0|x_0){\rm Res}_{z_2=\lambda_0}\tilde{p}(y;z_2|y_0)-\tilde{p}(y;z-\lambda_0|x_0){\rm
    Res}_{z_2=\lambda_0}\tilde{p}(x;z_2|y_0)\bigg\},
\end{split}
\end{equation}
which is the sought-for generalization of \eqref{invLapf>0}. We observe that
the double-residue term drops out due to the antisymmetrization procedure.
\end{widetext}

\subsection{Results}\label{SecResults}

\begin{figure}[tbp]
\centering
\includegraphics[width=9cm]{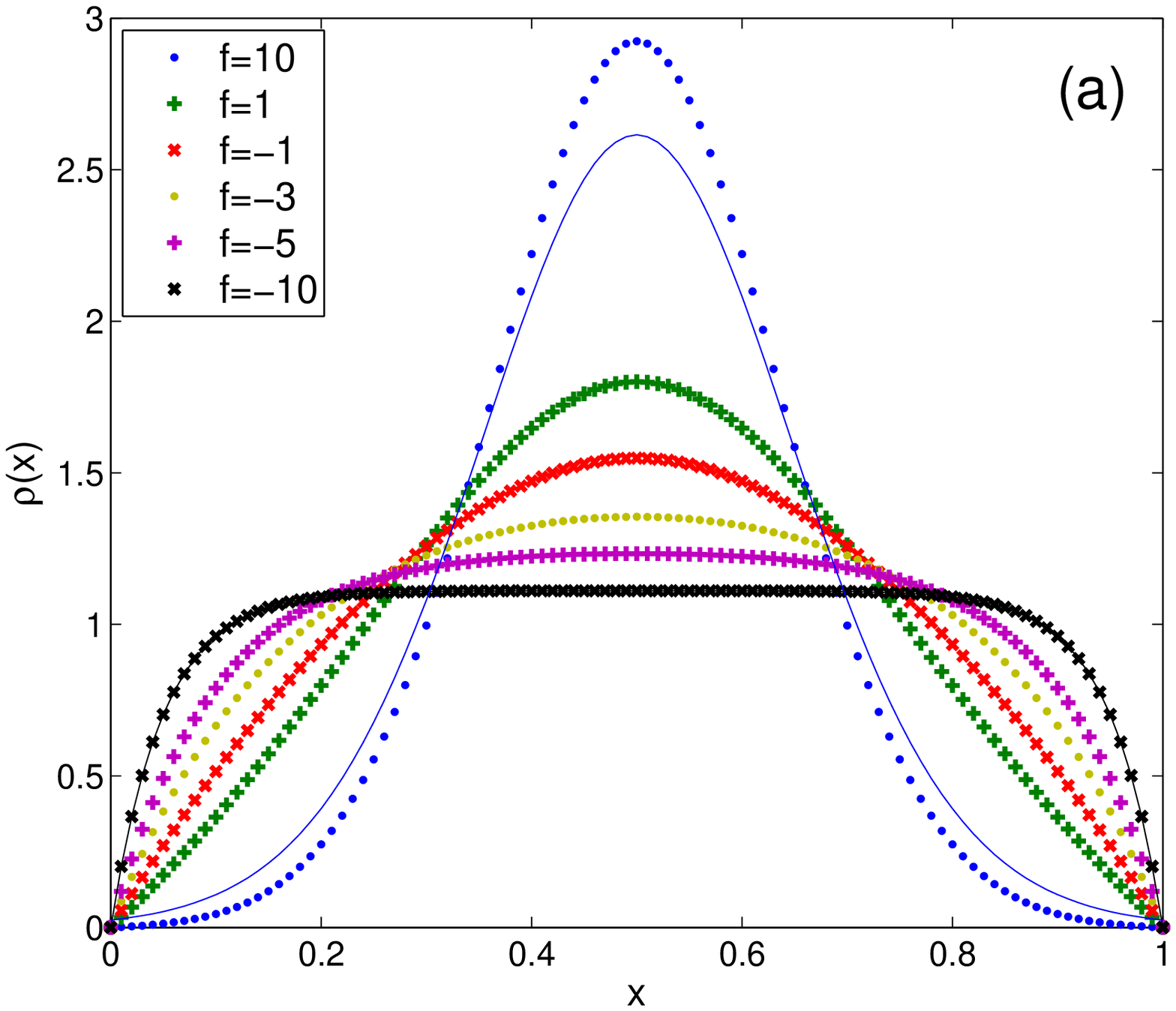}
\includegraphics[width=9cm]{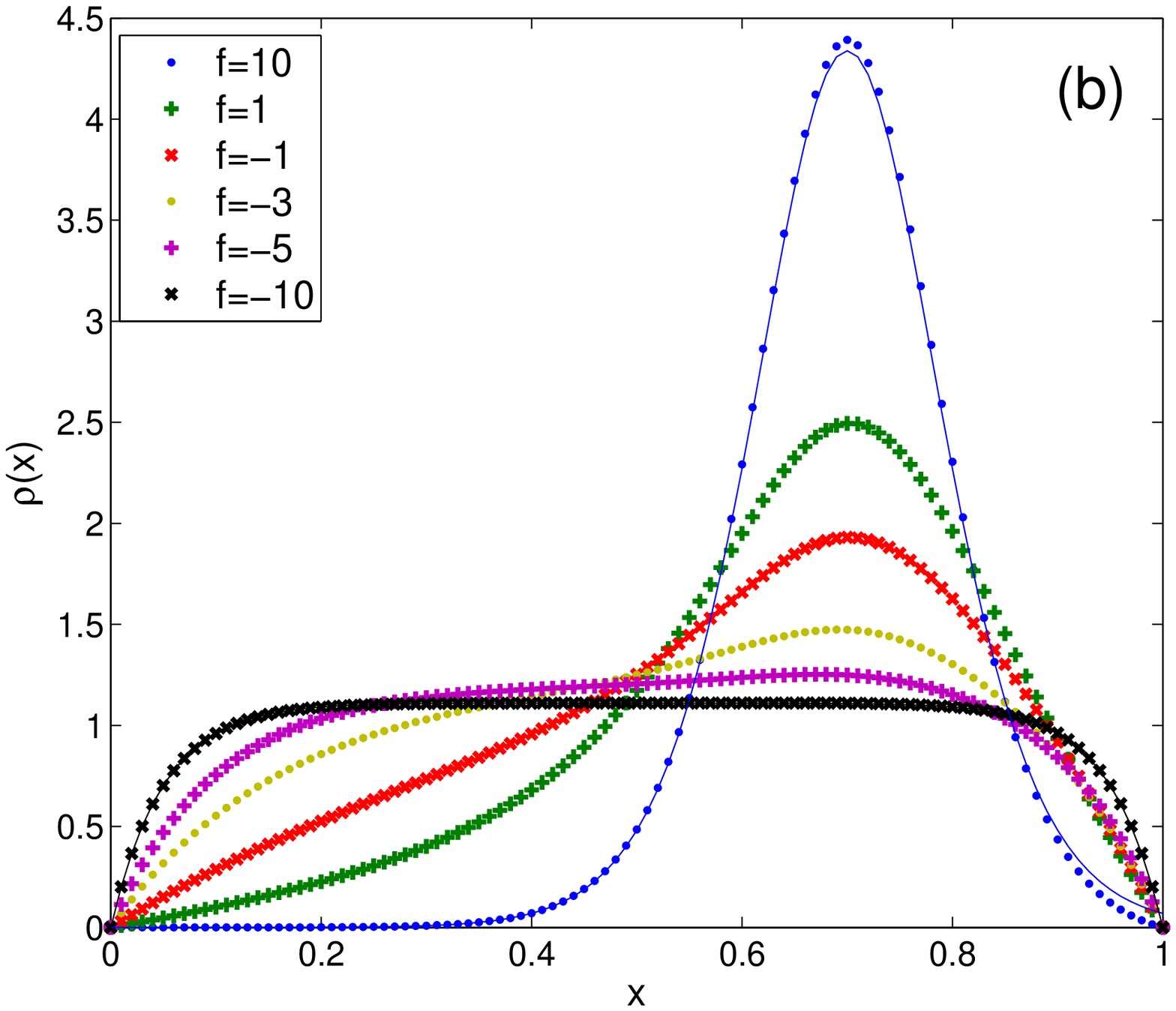}
\caption{Probability density for the coalescence position
$\rho(x|x_0,y_0)$ as a function of the position $x$ for several
values of the dimensionless force $f$. The initial positions of the
walkers $x_0$ and $y_0$ were $x_0=0,\,y_0=1$ (a) and
$x_0=0.5,\,y_0=0.9$ (b). The full lines for cases $f=10$ and $f=-10$
correspond to analytical results in Eqs. \eqref{abco_freefall} and
\eqref{boundary_limit}, respectively, for given initial conditions.}
\label{abco_fig}
\end{figure}

\begin{figure}
\includegraphics[width=9cm]{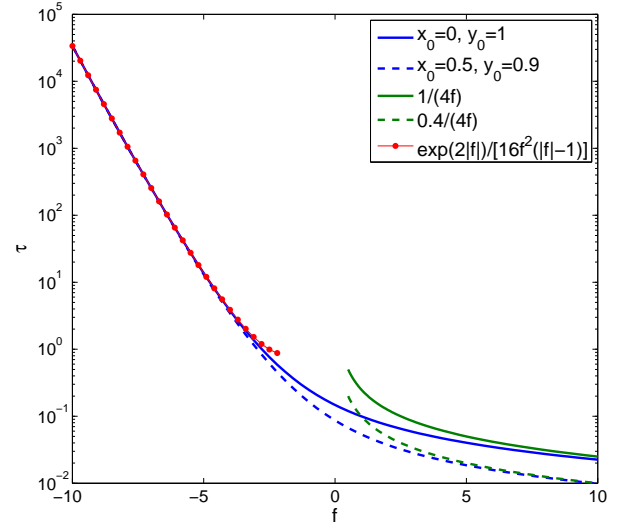}
\caption{The mean coalescence time $\tau(x_0,y_0)$ as a function of
the dimensionless force $f$ for two different initial conditions
$x_0=0,\,y_0=1$ (full line) and $x_0=0.5,\,y_0=0.9$ (dashed line).
The asymptotic analytical results for large positive, i.e.~``free
fall" case with $\tau=(y_0-x_0)/(4f)$, and negative, i.e.~large
barrier case of Eq.~\eqref{MFPT_limit}, forces are also shown for
comparison.} \label{mfpt_fig}
\end{figure}

Eqs.~\eqref{invLapf>0} and \eqref{invLapf<0} for $f>0$ and $f<0$,
respectively, together with the single-walker Green's function
\eqref{1partGF} and the identities for $\rho(x|x_0,y_0)$
\eqref{abco} and $\tau(x_0,y_0)$ \eqref{mfpt} were implemented in
{\tt \small MATHEMATICA} and evaluated. The results are shown in
Figs.~\ref{abco_fig} and \ref{mfpt_fig} depicting the coalescence
position probability density for several values of $f$ and the mean
coalescence time as a function of $f$, respectively. Both quantities
are shown for two different initial conditions $x_0=0,\,y_0=1$ (the
two walkers start out right at the boundaries) and
$x_0=0.5,\,y_0=0.9$ (a generic initial condition). Further results
for the coalescence time probability density $\pi(t|x_0,y_0)$
\eqref{Eqpi} are presented in later sections.

The results for the coalescence position in Fig.~\ref{abco_fig} show
for both initial conditions a clear crossover from a peaked form of
the probability density for large positive force (the case of an
almost ``free fall" into the potential well where the boundary
conditions have negligible influence on the dynamics, studied in
detail in Ref.~\cite{hame} and discussed below) to a very flat
probability density in the case of large negative force
corresponding to a high barrier. The flatness in the latter case can
be understood from a simple Arrhenius-like model, in which the
probability of the walker to be at a place $x$ is proportional to a
Boltzmann weight, i.e., $\exp[-\beta\phi(x)]$, where
$\phi(x)=-\int^x F(x')dx'$ is the free energy corresponding to the
force $F(x)= \pm f$. Since we are now dealing with two walkers, the
probability of both of them being simultaneously at the coalescence
position is given by the product of the Boltzmann weights,
$\exp(-\beta[\phi_L(x)+\phi_R(x)])$, which is a position-independent
constant due to the cancelation of the position-dependence of the
two opposite linear potentials. This simple picture breaks down
close to the boundaries but otherwise is sufficient to grasp the
observed behavior.

The characteristic (mean) coalescence times in Fig.~\ref{mfpt_fig}
cross over from the ``free fall" behavior for $f\gg 1$, proportional
to the inverse of the force $\tau\simeq 1/f$ (using the natural
boundary condition in the calculation of Ref.~\cite{hame} gives
$\tau=(y_0-x_0)/(4f)$, cf.\ Eq.~(9) therein) to the thermal
Arrhenius/Kramers-like barrier crossing proportional to the
exponential of the barrier height $\tau\simeq \exp(2|f|)$ for $f\ll
-1$. Thus, all the results are plausible and can be qualitatively
rationalized based on simple physical arguments.

In the rest of this section we focus on a more detailed study of the
two limiting cases, the almost ``free fall" $f\gg 1$ and the large
barrier $f\ll -1$. For these limiting cases we obtain analytical
results from relatively simple assumptions which compare
quantitatively well with the full solution. We start with the ``free
fall" (ff) case where we assume that the large drift towards
coalescence dominates the dynamics so that the reflecting boundary
conditions can be safely neglected since the typical
realizations/trajectories of the stochastic process never reach
them. The validity of this assumption depends on the initial
conditions and we expect it to be good enough for the walkers
starting from initial positions $x_0,\,y_0$ satisfying $\mathrm{min}
(x_0, 1-y_0)\gtrsim 1/f$, i.e.\ far enough from the boundaries.
This, indeed, turns out to be the case, see below and compare the
corresponding results in Figs.~\ref{abco_fig}a and \ref{abco_fig}b.

If the reflecting boundary conditions are neglected the 2-walker
Fokker-Planck equation \eqref{redfpe} can be solved by separation of
variables. In particular, the reformulation of Eq.~\eqref{redfpe}
together with conditions \eqref{EqVicious} and \eqref{init_cond} in
terms of the center-of-mass $(x+y)/2$ and relative $y-x$ variables
leads to a separable problem which can be easily solved since the
center-of-mass coordinate (cms) just performs a free diffusion while
the relative coordinate (rel) satisfies equations analogous to those
in Ref.~\cite{hame} (Eq. (7) with c=0 therein). Following a
derivation similar to that in Sec.~\ref{general} we arrive at (for
details the reader is referred to an upcoming publication
\cite{nome})
\begin{equation}
\varrho_{\rm ff}(x;t|x_0,y_0)= P_{\rm
cms}\big(\frac{x+y}{2};t\big|\frac{x_0+y_0}{2}\big)\pi_{\rm
rel}(t|y_0-x_0),
\end{equation}
with
\begin{equation}
P_{\rm cms}(u;t\big|u_0)=\frac{1}{\sqrt{2\pi
t}}\exp\left(-\frac{(u-u_0)^2}{2 t}\right)
\end{equation}
being the free diffusion propagator of the center-of-mass coordinate
(see Refs.~\cite{aslangul,tobiasbob}) and
\begin{equation}
\pi_{\rm rel}(t|y_0-x_0)=\frac{y_0-x_0}{\sqrt{8\pi
t^3}}\exp\left(-\frac{(y_0-x_0-4ft)^2}{8 t}\right)
\end{equation}
the first passage time probability density for the relative
coordinate to reach the origin (compare with Eq.~(8) in
Ref.~\cite{hame}). Using
$\rho(x|x_0,y_0)=\int_0^{\infty}dt\,\varrho(x;t|x_0,y_0)$ and the
identity $\int_0^{\infty}dt\exp(-a^2/(2t)-2 b^2 t)/t^2=4b
\mathrm{K}_1(2ab)/a$ for $a,b>0$ ($\mathrm{K}_n(x)$ is the modified
Bessel function of the second kind of order $n$) we finally obtain
for the coalescence position probability density in the ``free fall"
limit
\begin{equation}\label{abco_freefall}
\begin{split}
&\rho_{\rm ff}(x|x_0,y_0)=
\frac{f(y_0-x_0)\exp[f(y_0-x_0)]\,\mathrm{K}_1(2 f r)}{\pi\, r},\\
&\text{with}\\
&r\equiv\sqrt{(x-(x_0+y_0)/2)^2+(y_0-x_0)^2/4}.
\end{split}
\end{equation}
This function is plotted in Fig.~\ref{abco_fig}a,b for the two
different initial conditions for force $f=10$. We can see a rather
good agreement between the asymptotic formula \eqref{abco_freefall}
and the full result in Fig.~\ref{abco_fig}b. The situation is much
worse in Fig.~\ref{abco_fig}a although even there the correspondence
is qualitatively quite acceptable. As mentioned above, the reason
for the success or failure of the approximation is determined by the
initial conditions. Indeed, Fig.~\ref{abco_fig}b corresponds to $f
\mathrm{min} (x_0, 1-y_0)=f(1-y_0)=1$ where the approximation is
expected to become valid while in Fig.~\ref{abco_fig}a the walkers
start out right at the boundaries and only an extremely high value
of the force could prohibit the walkers from occasionally bumping
into the boundaries, especially at the very beginning. Thus, in the
case with $x_0=0,\,y_0=1$ the above approximation is expected to
become quantitatively accurate only for very high values of $f$ ---
numerical estimates reveal that an agreement comparable with that of
Fig.~\ref{abco_fig}b is not achieved until about $f\gtrsim 40$. This
supports a heuristic guess that the accuracy of the asymptotic
analytic expression crosses over from
$\exp[-f\mathrm{min}(x_0,1-y_0)]$ to $1/f$ when the minimum equals
zero.

Now, we turn to the opposite limit of large barrier, i.e.\ the
$f\ll-1$ case. In particular, we want to derive asymptotic
expressions for the characteristic coalescence time (which is
identified from the results of the full theory as
$\tau=1/|\Lambda_0|$, cf.~the end of Sec.~\ref{SecCalgt} and below)
and the coalescence position probability density. Clearly, this
limit is dominated by the lowest eigenvalue and eigenfunction of the
2-walker problem which means by the two lowest eigenvalues and
eigenfunctions of the auxiliary 1-walker problem. Thus we can write,
under assumptions of large barrier (lb) and generic initial
conditions (to be specified in more detail below), for $P_{\rm
lb}(x,y;t|x_0,y_0)$,
\begin{eqnarray}
\nonumber
P_{\rm lb}(x,y;t|x_0,y_0)&\simeq& e^{f(x-y)}e^{-f(x_0-y_0)}e^{(\lambda_0+\lambda_1)t}\\
\nonumber
&&\hspace*{-0.8cm}\times [\psi_0(x)\psi_1(y)-\psi_0(y)\psi_1(x)]\\
&&\hspace*{-0.8cm}\times
[\psi_0(x_0)\psi_1(y_0)-\psi_0(y_0)\psi_1(x_0)], \label{1mode}
\end{eqnarray}
where $\lambda_{0,1}\simeq\pm 4f^2e^{-|f|}$ are the lowest
eigenvalues satisfying Eq.~\eqref{seculareq1} and $\psi_{0,1}(x)$
are the corresponding eigenfunctions given by (these are exact
expressions for all $f<-2$)
\begin{equation}
\begin{split}
    \psi_0(x) &=\sqrt{\frac{2\lambda_0}{\lambda_0+2|f|}} \cosh[\sqrt{f^2+\lambda_0}(x-1/2)],\\
    \psi_1(x) &=\sqrt{\frac{2|\lambda_1|}{\lambda_1+2|f|}}
    \sinh[\sqrt{f^2+\lambda_1}(x-1/2)]\ .
\end{split}
\end{equation}
Using these expressions we can study the dependence on the initial
conditions and clarify the regime in which the assumption about
the dominance of the lowest eigenmode is valid. Utilizing that in
the limit $f\ll-1$ the eigenvalues satisfy
$|\lambda_1|\simeq\lambda_0\ll|f|$ we obtain
\begin{eqnarray}
\nonumber
&&e^{-f(x_0-y_0)}[\psi_0(x_0)\psi_1(y_0)-\psi_0(y_0)\psi_1(x_0)]\\
\nonumber
&&\hspace*{1.2cm}\simeq\frac{\lambda_0}{|f|}e^{-f(x_0-y_0)} \sinh[|f|(y_0-x_0)]\\
&&\hspace*{1.2cm}=\frac{\lambda_0}{2|f|}(1-\exp[-2|f|(y_0-x_0)])\ .
\end{eqnarray}
Since by assumption $y_0>x_0$ we see that the evolution depends only
exponentially weakly on the initial conditions so that for
$y_0-x_0\gg1/|f|$ the evolution is essentially independent of the
initial conditions as expected in the high barrier limit. Indeed,
the above condition just says that the walkers should start out well
separated so that the barrier between them is still large (in
dimensionless units). In such a case the dynamics is independent of
the detailed initial condition or, more precisely, it depends on it
only exponentially weakly which can safely be neglected. This is the
regime in which the lowest eigenmode theory of Eq.~\eqref{1mode} is
sufficient as we will demonstrate below.

If we calculate $\varrho(x;t|x_0,y_0)$ from
Eq.~\eqref{boundary_def2} in the limit $f\ll-1,\,y_0-x_0\gg1/|f|$ we
obtain
\begin{equation}\label{varrho}
\begin{split}
    \varrho_{\rm lb}(x;t|x_0,y_0)&\simeq
    \frac{\lambda_0}{|f|}
    e^{(\lambda_0+\lambda_1)t}[\psi_0(x)\psi'_1(x)-\psi_1(x)\psi'_0(x)]\\
    &\hspace*{-1.2cm}\simeq \frac{\lambda_0^2}{|f|}e^{(\lambda_0+\lambda_1)t}\Big\{1-\frac{\lambda_0}{2
    f^2}\cosh[2|f|(x-1/2)]\Big\}.
\end{split}
\end{equation}
Now integrating over time and taking into account that
$\lambda_0\simeq 4f^2e^{-|f|}$ and
$\lambda_0+\lambda_1\equiv\Lambda_0\simeq -16f^2(|f|-1)e^{-2|f|}<0$
we get for $\rho(x|x_0,y_0)=\int_0^{\infty}dt\,\varrho(x;t|x_0,y_0)$
\begin{eqnarray}
\nonumber \rho_{\rm
lb}(x|x_0,y_0)&\simeq&\frac{|f|}{|f|-1}\Big(1-2e^{-|f|}\cosh[2|f|(x-1/2)]
\Big)\\
&=&\frac{1}{1-\frac{1}{|f|}}\Big(1-e^{-2|f|x}-e^{-2|f|(1-x)}\Big).
\label{boundary_limit}
\end{eqnarray}
This clarifies that $\rho_{\rm lb}(x|x_0,y_0)$ is properly
normalized to one within exponential precision, $\int_0^1
dx\,\rho_{\rm lb}(x|x_0,y_0)=
1+\mathcal{O}\left(e^{-2|f|}\right)\simeq 1$ , which finally proves
the self-consistency of the lowest eigenmode approximation. The
curves for the case $f=-10$ in Fig.~\ref{abco_fig}a,b calculated by
the full theory are practically indistinguishable from that given by
the approximate expression \eqref{boundary_limit} (which is an
explicit illustration of the initial-condition independence). In a
straightforward manner it also follows that the mean coalescence
time is given by the inverse lowest eigenvalue $1/|\Lambda_0|$ since
for $\varrho_{\rm lb}(x;t|x_0,y_0)$ in the separable form of
Eq.~\eqref{varrho} and due to the above normalization condition one
immediately gets
\begin{eqnarray}
\nonumber
\tau_{\rm lb}(x_0,y_0)&\simeq& \int_0^{\infty}dt\, t\int_0^1 dx\,\varrho_{\rm lb}(x;t|x_0,y_0)\\
\nonumber
&=&\int_0^{\infty}dt\, t e^{-|\Lambda_0|t}|\Lambda_0|\int_0^1 dx\,
\rho_{\rm lb}(x|x_0,y_0)\\
&\simeq&\frac{1}{|\Lambda_0|}\simeq\frac{e^{2|f|}}{16f^2(|f|-1)},
\label{MFPT_limit}
\end{eqnarray}
independent of the initial conditions. All approximate equalities
hold up to exponentially small corrections of order $e^{-2|f|}$,
which are negligible for $f\ll-1$.

\subsection{Summary}

We have in Secs.~\ref{SecMEFP}, \ref{SecSolutionFP}, and
\ref{SecBubbleCoa} set up an approximate Fokker-Planck equation
scheme for the full problem of two interfaces moving in a block
DNA-stretch with a barrier region separating two soft zones. While
the full problem can be viewed as two discrete random walkers in
different potentials with an imposed vicious boundary condition, the
approximate Fokker-Planck equation describes two continuous random
walkers in opposite linear potentials keeping the imposed
viciousness condition. The four assumptions leading to the
Fokker-Planck equation were introduced in Sec.~\ref{SecMEFP} and
their validity will be discussed favorably in
Sec.~\ref{SecComparison} when we compare the Fokker-Planck results
with the direct evaluation of the full discrete problem using the
master equation approach presented in the next
section.\\

The main outcome of the Fokker-Planck approach are the general
results for the coalescence time density $\pi(t)$ (examples will be
shown in Sec.~\ref{SecComparison}), and the numerical and analytic
expressions for the mean meeting time $\tau$ (shown in
Fig.~\ref{mfpt_fig}) as well as the probability density $\rho$ for
the meeting position (shown in Fig.~\ref{abco_fig}). All results are
expressed through the dimensionless force $f$, which in terms of the
parameters from an experimental setup reads
\begin{equation}
f=\frac{N(u_b-1)}{u_b+1},
\end{equation}
making comparison with values obtained from experiments
straightforward.

Finally, a remark on why one should consider the continuous approach
over the complete, discrete master equation approach is in order.
Namely, for DNA-stretches of length $N$ the discrete master equation
approach involves diagonalization of matrices of the order
$N^2\times N^2$, setting computational limitations on $N$ and a new
diagonalization is needed for each parameter set. The Fokker-Planck
approach may therefore provide additional insight for very long DNA
stretches as we showed here by discussing the physical quantities of
meeting position and meeting time.

\section{Complete discrete approach: the master equation}\label{SecME}

In this section, we develop a master equation framework for the
bubble coalescence. In contrast to the previous treatment, we
explicitly allow the soft zones to zip close from the two ends of
the barrier region. It will turn out that in some cases this only
has a minor effect. A detailed comparison with the continuum
Fokker-Planck equation approximation is shown in the next section.

We consider the same segment of double-stranded DNA with
$M=N_L+N+N_R$ internal base-pairs, clamped open at both ends
according to Fig.~\ref{fig:model}. However, in contrast to the
approximations imposed in the Fokker-Planck approximation, we now
allow for explicit closure of the soft regions, necessitating the
consideration of a sequence-dependence of the local DNA stability,
in contrast to the previous discussion, where we assumed that the
two bubble domains are always remaining open.

Note that in this section our notation differs from the scheme
introduced above, first, in order to keep the notation of this
section consistent with previous references on the same method
\cite{JPC,tobiasprl,tobiasbj,tobiasrc,tobiaspre}, and second, to be
able to incorporate zipping/unzipping of base pairs in the soft
zones, as well. We denote by $x_L=X+N_L$ ($x_R=Y+N_L$) the position
of the rightmost (left-most) open basepair in the left (right) open
region, see Fig.~\ref{fig:model}, where $x_{L,R}\in [0,M]$. The
positions $x_L$ and $x_R$ of the two zipper forks are stochastic
variables and the aim is to understand how these variables evolve in
time without taking the continuum limit and using the approximations
introduced in Sec.~\ref{SecMEFP}. We note that an equivalent set of
variables are $x_L$ and the clamp size $m$, that are related through
  \begin{equation}
m=x_R-x_L.
  \end{equation}
In the master equation formulation below we will use $x_L$ and $m$
as the dynamic variables, and for completeness we state the
transition rates, Eqs.~(\ref{Eqreflecting1}), (\ref{Eqreflecting2}),
(\ref{eq:t_L_plus}), (\ref{eq:t_R_minus}), and (\ref{eq:t_L_minus}),
(\ref{eq:t_R_plus}) expressed in the new variables: The reflecting
boundary conditions are
  \begin{equation}
t_L^-(x_L=0,m)=t_R^+(x_L,m=M-x_L)=0.\label{eq:reflecting2b}
  \end{equation}
Once the clamp is completely unzipped, i.e. the state $m=0$ is
reached, we assume that the clamp will not be able to reform for a
long time, and we impose the absorbing conditions
  \begin{equation}
t_L^-(x_L,m=0)=t_R^+(x_L,m=0)=0.\label{eq:absorbingb}
  \end{equation}
The transition rates at the interior of the DNA stretch is
  \begin{eqnarray}
t^-_L(x_L,m)&=&\frac{1}{2} \mathcal{K}(x_L),\label{eq:t_L_plusB}\\
t^+_R(x_L,m)&=&\frac{1}{2}\mathcal{K}(M-m-x_L),\label{eq:t_R_minusB}\\
t_L^+(x_L,m)&=&\frac{1}{2}\mathcal{K}(x_L+1)u(x_L+1)s(m),\label{eq:t_L_minusB}\\
t_R^-(x_L,m)&=&\frac{1}{2}\mathcal{K}(M-m-x_L+1) u(x_R-1)\nonumber\\
& &\times s(M-m-x_L),\label{eq:t_R_plusB}
  \end{eqnarray}
where $x_R=x_L+m+1$, $s(q)=\{(q+1)/(q+2)\}^c$ and $\mathcal{K}(q)=k
q^{-\mu}$. The properties of the four transfer coefficients above
are summarized in Table \ref{table:transfer_coeff}.

\begin{table}
\vspace{0.2cm}
\begin{tabular}{@{}|l|r|r|r|c|}
  \hline
&$\Delta x_L$  & $\Delta x_R$ & $\Delta m$ & Eq.\\
  \hline
$t_L^+(x_L,m)$  & $1$    & $0$  & $-1$ & \eqref{eq:t_L_plusB}\\
   \hline
$t_R^-(x_L,m)$    & $0$  & $-1$  &  $-1$ & \eqref{eq:t_R_minusB} \\
   \hline
$t_L^-(x_L,m)$  & $-1$    & $0$  & $1$ & \eqref{eq:t_L_minusB}\\
   \hline
$t_R^+(x_L,m)$    & $0$  & $1$  &  $1$ & \eqref{eq:t_R_plusB}\\
   \hline
  \end{tabular}
\caption{\label{table:transfer_coeff} Properties of the
transfer coefficients. The quantity
$\Delta x_L$ ($\Delta x_R)$ denotes the change in fork position
$x_L$ ($x_R$) under the action of the transfer coefficients. Similarly
$\Delta m$ denotes the change in clamp size $m$.}
\end{table}

Denote by $P(x_L,m,t|x_L',m')$ the conditional probability to find
the system in state $(x_L,m)$ at time $t$, given the initial
condition $(x_L',m')$ at time $t=0$. With the short-hand notation
$P(x_L,m,t)=P(x_L,m,t|x_L',m')$ the dynamics is described by the
master equation
\begin{eqnarray}
& &\hspace*{-0.8cm}\frac{\partial}{\partial t}P(x_L,m,t)=\nonumber\\
& &t_L^+(x_L-1,m+1)P(x_L-1,m+1,t)\nonumber\\
& &+ t_L^-(x_L+1,m-1)P(x_L+1,m-1,t)\nonumber\\
& &-\left[ t_L^+(x_L,m)+t_L^-(x_L,m)\right] P(x_L,m,t)\nonumber\\
& &+t_R^+(x_L,m-1)P(x_L,m-1,t)\nonumber\\
& &+ t_R^-(x_L,m+1)P(x_L,m+1,t)\nonumber\\
& &-\left[ t_R^+(x_L,m)+t_R^-(x_L,m)\right]
P(x_L,m,t).\label{eq:master_eq1}
  \end{eqnarray}
This equation states that the probability for the clamp size can
change in 8 different ways: the terms with a plus-sign correspond to
jumps {\em to} the state $\{x_L,m\}$, and the terms with a minus-sign
correspond to jumps {\em from} the state $\{x_L,m\}$.

A standard approach to the master equation (\ref{eq:master_eq1}) is the
spectral decomposition \cite{van_Kampen,Risken}
  \begin{equation}
P(x_L,m,t)=\sum_p c_p(x_L',m')Q_p(x_L,m)\exp(-\eta_p
t),\label{eq:spectral_decomp}
\end{equation}
in terms of the eigenvalues $\eta_p$ and eigenvectors
$Q_p(x_L,m)$. The expansion coefficients $c_p(x_L',m')$ are
obtained from the initial condition. As in the previous section,
we will assume the system initially to be in the state where all
bps in the soft zone are broken and all bps in the barrier region are
closed.

The eigenvalue equation corresponding to Eq. (\ref{eq:master_eq1})
becomes
 \begin{eqnarray}
& &t_L^+(x_L-1,m+1)Q_p(x_L-1,m,t)\nonumber\\
& &+ t_L^-(x_L+1,m-1)Q_p(x_L+1,m-1,t)\nonumber\\
& &-\left[ t_L^+(x_L,m)+t_L^-(x_L,m)\right] Q_p(x_L,m,t)\nonumber\\
& &+t_R^+(x_L,m-1)Q_p(x_L,m-1,t)\nonumber\\
& &+ t_R^-(x_L,m+1)Q_p(x_L,m+1,t)\nonumber\\
& &-\left[ t_R^+(x_L,m)+t_R^-(x_L,m)\right] Q_p(x_L,m,t)\nonumber\\
& &\hspace*{0.8cm}=-\eta_p Q_p(x_L,m,t).\label{eq:eigenvalue_eq1}
  \end{eqnarray}
The eigenvectors satisfy the orthogonality relation
\cite{van_Kampen}
  \begin{equation}
\sum_{m=1}^M \sum_{x_L=0}^{M-m} \frac{Q_p(x_L,m)
Q_{p'}(x_L,m)}{P^{\rm eq}_r(x_L,m)}=\delta_{p p'},
\label{eq:ortho_relation}
  \end{equation}
where
\begin{equation}
P_r^{\rm eq}(x_L,m)=\mathscr{Z}(x_L,m)/\mathscr{Z},
\end{equation}
with
\begin{equation}
\mathscr{Z}=\sum_{x_L,m}Z(x_L,m),
\end{equation}
being the partition coefficient in the variable $x_L$ and $m$
(neglecting a common $\sigma_0^2$-factor, where $\sigma_0$ is the
bubble initiation parameter), and where
\begin{eqnarray}
\mathscr{Z}(x_L,m)&=&\prod_{\tilde{x}_L=0}^{x_L} u(\tilde{x}_L)
\prod_{\tilde{x}_R=m+x_L+1}^{M+1} u(\tilde{x}_R)\nonumber\\
& &\hspace{-1cm}\times (1+x_L)^{-c}(M-m-x_L+1)^{-c}, \label{eq:Z}
\end{eqnarray}
see Eqs.~\eqref{EqZLX}-\eqref{EqFullPartition}. From the eigenvalues
and eigenvectors of Eq.~(\ref{eq:eigenvalue_eq1}) any quantity of
interest may be constructed. In the following subsection we
calculate the coalescence time density. How to set up the master
equation for numerical purposes is presented in detail in
App.~\ref{AppMEnumerical}. Alternatively, the master equation
\eqref{eq:master_eq1} can be solved by direct stochastic simulations
such as, e.g.~the Gillespie algorithm introduced in
App.~\ref{AppGillespie} and used in Sec.~\ref{SecComparison}.

\subsection{Coalescence time density}

The (survival) probability that the absorbing boundary
at $m=0$ has not yet been reached up to time $t$ is
\begin{equation}
\mathscr{S}(x_L',m',t)=\sum_{m=1}^M\sum_{x_L=0}^{M-m}
P(x_L,m,t|x_L',m').
\end{equation}

The probability that the absorbing boundary is reached within the
time interval $[t,t+dt]$ (namely, the coalescence time density corresponding
to the first passage problem) is
\begin{eqnarray}
\nonumber
\rho (x_L',m',t)dt&=&\mathscr{S}(x_L',m',t)-\mathscr{S}(x_L',m',t+dt)\\
&=&-\left(\frac{\partial}{\partial t}\mathscr{S}(x_L',m',t)\right)dt.
\label{eq:fptd1}
\end{eqnarray}
This expression is positive, as $\mathscr{S}$ is decreasing with time.
To express the coalescence time density in terms of the eigenvalues
and eigenfunctions, we introduce the eigenmode expansion
(\ref{eq:spectral_decomp}) into equation (\ref{eq:fptd1}), yielding
\begin{equation}
\rho (x_L',m',t)=\sum_p\eta_p c_p(x_L',m') \exp(-\eta_pt),
\label{eq:fptd2}
\end{equation}
with coefficients
\begin{equation}
c_p(x_L',m')=\frac{Q_p(x_L',m')}{P_r^{\rm eq}(x_L',m')}\sum_{m=1}^M
\sum_{x_L=0}^{M-m}Q_p(x_L,m).
\label{eq:c_p}
\end{equation}
We have above made use of the orthonormality relation
(\ref{eq:ortho_relation}) in order to express $c_p(x_L',m')$ in
terms of the initial probability density $P(x_L,m,0|x_L',m')$, and
used the fact that this general initial condition takes the
explicit form $P(x_L,m,0|x_L',m')=\delta_{x_L,x_L'}\delta_{m,m'}$.
Eq. (\ref{eq:fptd2}) is the discrete counterpart of the continuous
result derived in Ref.~\cite{Gardiner}, and expresses the coalescence
time density (for any given initial condition, specified by $m'$
and $x_L'$) in terms of the eigenvalues and eigenvectors of Eq.
(\ref{eq:eigenvalue_eq1}).

\section{Comparison between the full master equation and the Fokker-Planck
approximation}
\label{SecComparison}

In this section, we investigate the validity of the assumptions
presented in Sec.~\ref{SecMEFP} leading to the Fokker-Planck
continuum approximation. This is done by comparing the results for
the coalescence time densities, $\pi(t)$, with the results obtained
from the full discrete master equation approach. In the next
section, Sec.~\ref{SecBio}, we discuss the relevance for biological
experiments.

In all examples below the two walkers move between $[0,N]$ in the
Fokker-Planck description and between $[-N_L,N+N_R]$ in the master
equation setup. That is, in the master equation approach we
explicitly allow zipping of base pairs in the two soft zones. As
initial conditions we use $X_0=0$ and $Y_0=N$ throughout this
section.

\subsection{The continuum approximation}

The continuum assumption ({\it iv}) implies that the inherently
discrete nature of the DNA structure - both in terms of stacking and
hydrogen bonds - can be approximated by the diffusive behavior of
two continuous variables. To get the Fokker-Planck description one
has to consider the limit $a\rightarrow 0$ with $a$ being the length
between effective bonds in the base pairs. In practice this limit is
obtained by
\begin{equation*}
    \frac{\text{bond distance}}{\text{total segment length}}\rightarrow 0,
\end{equation*}
i.e. by increasing the width of the barrier region. The question
is addressed in a setup with open soft zones, i.e. $ u_{s}\gg 1$,
and varying barrier lengths, $N$. To have perfectly reflecting
boundary conditions we set $N_L=N_R=0$ in the master equation
setup.

\begin{figure}
    \includegraphics[width=.5\textwidth]{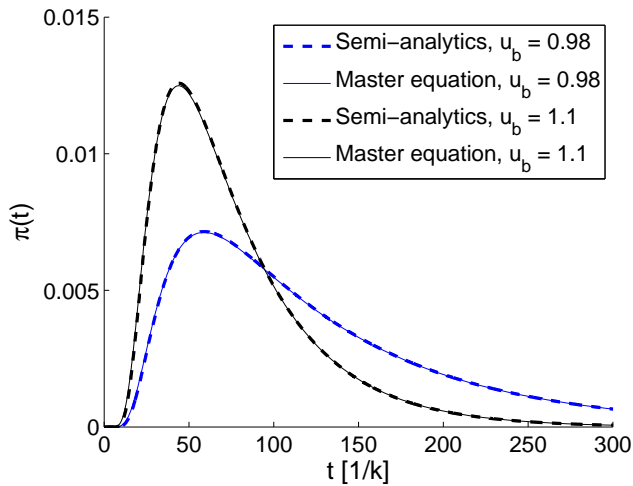}
   \caption{
Coalescence time probability density for barrier
   width $N=20$ and temperature below and above $T_b$, with $u_b=0.98$ and $u_b=1.1$, respectively.
   To exclude other effects, the lengths of the soft zones are zero, $N_L=N_R=0$,
   i.e., these are assumed to be always open, and $c=\mu=0$. }
   \label{FigContinuum}
  \end{figure}

Fig.~\ref{FigContinuum} shows that it requires a relatively small
number of base pairs ($\sim 20$) before the continuum approximation
is reasonable, independently of the temperature. Barrier regions of
this length are in principle accessible experimentally so the
continuum approximation appears to be well-justified.

\subsection{Open soft zones}

The first assumption, ({\it i}), states that the soft zones are
always open, i.e. that the random walkers are reflected at the
interfaces between the barrier region and the soft zones.

To eliminate other effects than the effect of the introduction of the
reflecting boundary conditions at the ends of the barrier region, we
consider a DNA stretch of length 25, so that the continuum
approximation is justified, and set $c=\mu=0$ in order to exclude
effects originating from the entropy factor and the hook exponent.
We compare this to the results from the full master equation including
the soft zones.

\begin{figure}[htbp]
    \centering
    \includegraphics[width=0.45\textwidth]{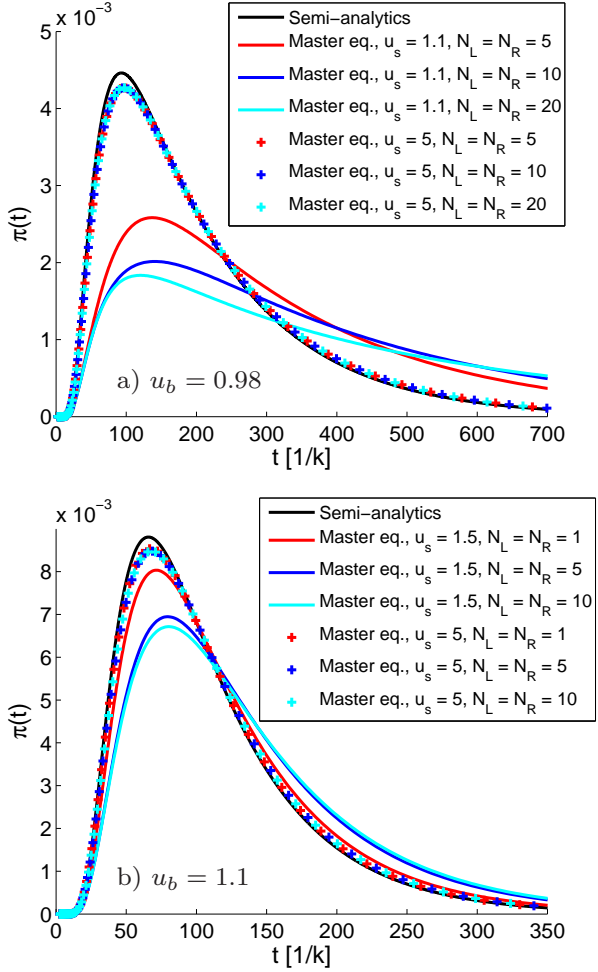}
    \caption{Coalescence time density for varying soft zone lengths and fixed length of the barrier $N=25$.
    Included are plots for $u_b=0.98$ (upper) and $u_b=1.1$ (lower), i.e. for temperatures
    below and above the melting temperature of the barrier, respectively. Furthermore, $c=\mu=0$.}
    \label{FigSoftZones}
  \end{figure}

Fig.~\ref{FigSoftZones} shows the coalescence time density $\pi(t)$
for varying lengths of the soft zones for temperatures above and
below the melting temperature of the barrier region. Apparently, the
length dependence is rather weak as long as the soft zones serve as
hard enough boundaries, i.e.\ for large enough $u_s\gtrsim 5$. For
smaller $u_s$ the soft-zone-length dependence is relevant as the two
bubble corners venture more frequently into the soft zones. This,
however, implies the breaking of the assumptions for the
applicability of our Fokker-Planck description as revealed in the
figure. Thus, the length of the soft zones itself is not important
if $u_s$ is large enough and $c=\mu=0$. Systematic assessment of
these conditions is given below.

\begin{figure*}[htbp]
    \includegraphics[width=0.9\textwidth]{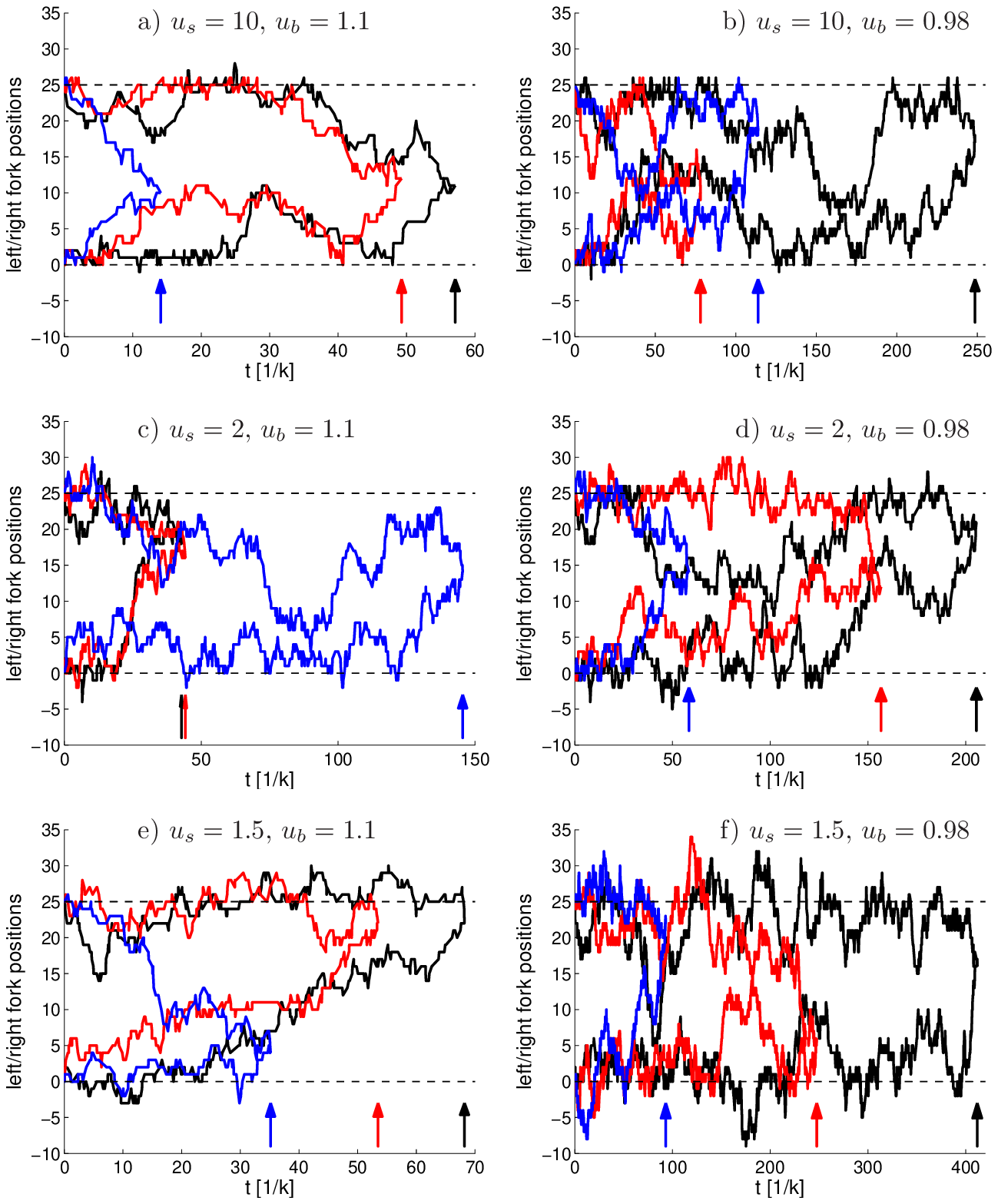}\label{Fig8all}
    \caption{ Single trajectories for the two zipper forks,
    based on the Gillespie (Monte Carlo) algorithm presented in App.~\ref{AppGillespie}.
    The values of $u_b$ and $u_s$ are stated in the figures. The other parameters are $N_L=N_R=20$, $N=25$, $c=\mu=0$.
    The dashed horizontal lines mark the boundaries between the barrier region and the soft zones, and the
    arrows indicate the ends of the sampled trajectories.}
    \label{FigTrajec}
  \end{figure*}

Fig.~\ref{FigVaryUb} shows $\pi(t)$ as a function of $u_s$ for
$N_L=N_R=20$, i.e. the soft zones are so long the two forks
essentially never reach the outer clamps. That this is indeed the
case can be qualitatively investigated using the Gillespie-scheme
presented in App.~\ref{AppGillespie}, giving access to real-time
trajectories of the two random walkers in a potential landscape
including both the barrier region and the soft zones. This is shown
in Fig.~\ref{FigTrajec} which confirms that excursions into the soft
zones are progressively suppressed with increasing $u_s$. In
Fig.~\ref{FigVaryUb} we have only included the case $u_b<1$, i.e.,
when the barrier region indeed acts as a barrier, and consequently
the effect of the soft zones is more pronounced. The discrepancies
between the Fokker-Planck and the master equation approach become
distinct for $u_s\simeq 1$ whereas for $u_s \gtrsim 5$ the agreement
between the approaches becomes reasonable. Difference between $u_s$
and $u_b$ of this magnitude can indeed be achieved in realistic
experimental setups, as shown in the next section.

\begin{figure}[tbp]
\centering
\includegraphics[width=0.45\textwidth]{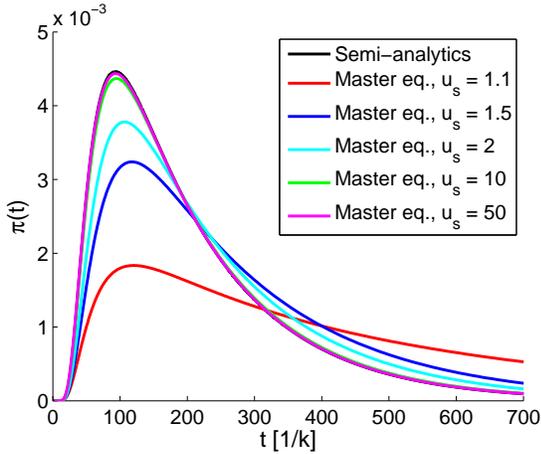}
\caption{Coalescence time density for $N=25$, $N_L=N_R=20$,
$u_b=0.98$ and different values of $u_s$. Furthermore, $c=\mu=0$.}
\label{FigVaryUb}
\end{figure}

\subsection{Loop Entropy and Hook factors}\label{SecEntropy}
The general rates defined in Sec.~\ref{SecRates} include both the
entropy loss factor and the hook factor. Both depend on the length
$q$ of the bubble, and for both their relative influence diminishes
for increasing bubble lengths. In the Fokker-Planck description both
factors are omitted, which are the assumptions {\it (ii)} and {\it
(iii)}. These assumptions are valid for long bubbles, which can be
obtained by having long soft zones and keeping the temperature far
above the melting temperature of the soft zones, i.e. $u_s\gg 1$.

\begin{figure}
    \centering
    \subfigure{\includegraphics[trim = 0mm 0mm 0mm 0mm, clip,width=.45\textwidth]{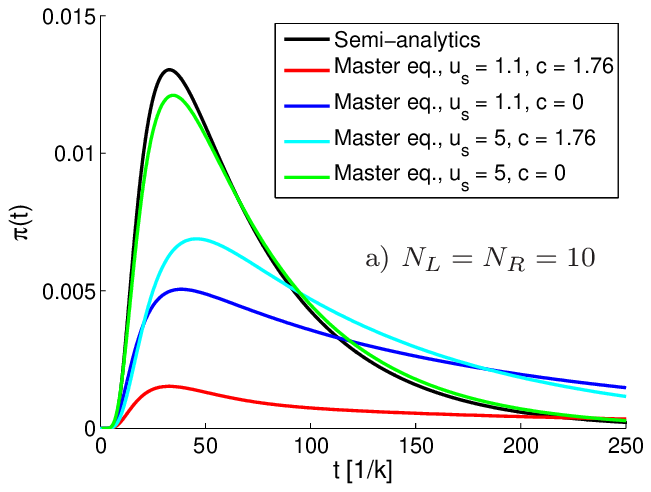}\label{FigLoop10}} \quad
    \subfigure{\includegraphics[trim = 0mm 0mm 0mm 0mm, clip,width=.45\textwidth]{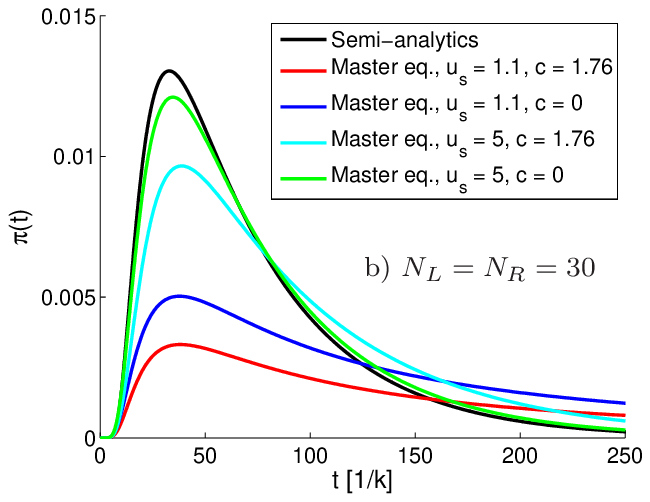}\label{FigLoop30}}
   \caption{Coalescence time density $\pi(t)$ for fixed barrier
   width $N=15$, and varying the length of the soft zones, $N_L=N_R=10$ in Fig.~\ref{FigLoop10} and
   $N_L=N_R=30$ in Fig.~\ref{FigLoop30}, respectively. The other parameters are $u_b=0.98$, $\mu=0$.}
   \label{FigLoops}
  \end{figure}

Fig.~\ref{FigLoops} studies the effect of the loop exponent $c$ on
the coalescence time density $\pi(t)$ for a fixed length of the
barrier region, $N=15$ and varying the lengths of the soft zones.
Even for soft zones of length $N_L=N_R=30$ there is a substantial
influence of the loop entropy factor, making long soft zones a
requirement in experimental realizations which should agree
reasonably with the Fokker-Planck approach.

\begin{figure}
     \includegraphics[width=.45\textwidth]{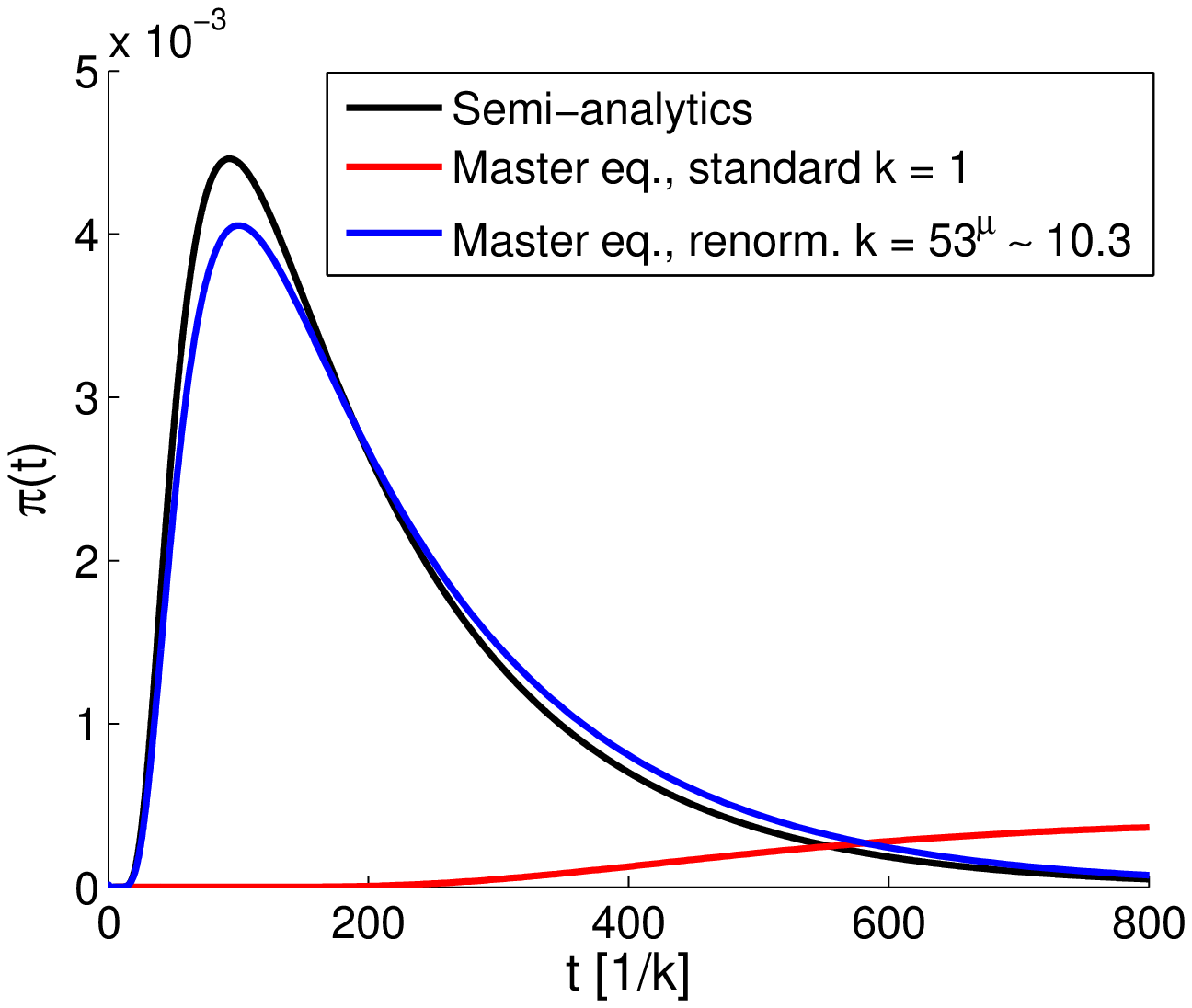}\label{Fig11}
   \caption{Coalescence time density $\pi(t)$ for non-vanishing hook exponent $\mu =0.588$, $N_L=N_R=50$ and
   $N=25$ for the barrier case, $u_b=0.98$. The results are shown with
  and without the renormalized rate constant $k$, where we have used $\tilde k = 53^{0.588}\simeq 10.3$.
 The other parameters are $u_s=10$  and $c=0$.} \label{FigLoopHookRates}
\end{figure}

The hook factor leads to a decrease in the rate constant $k$,
$k\rightarrow k q^{-\mu}$ where $q$ is the length of a given bubble,
so introducing the hook exponent $\mu > 0$ leads to a considerable
and bubble-length-dependent decrease of the transition rates, see
Fig.~\ref{FigLoopHookRates}. However, for sufficiently large
bubbles, the rate is roughly constant and introducing a renormalized
rate constant $\tilde{k}= k L^{\mu}$ can compensate for this effect.
Here $L$ is a characteristic bubble size. If the temperature is kept
well above the melting temperature of the soft zones, and the length
of the soft zones are much longer than the barrier, $L$ is well
approximated by the length of the soft zones, $L\simeq N_{L/R}$. In
Fig.~\ref{FigLoopHookRates} we illustrate the effect of a
renormalized rate constant (with $L=53$ being the best fit) together
with the standard rate coefficient $k=1$.

In conclusion, both the influence of the loop entropy factor and the
hook exponent can be eliminated by keeping the length of the soft
zones sufficiently long and using a renormalized value for the rate
constant $k$.

\section{Relevance for single molecule experiments}\label{SecBio}

Relevant for the separation of statistical weights are the
empirical relations \cite{krueger}
\begin{subequations}
\begin{eqnarray}
  T_m^{AT}&=& (355.55+7.95\ln[\mathrm{Na}^+])K \label{EqMeltAT}\\
  T_m^{GC}&=& (391.55+4.98\ln[\mathrm{Na}^+])K,\label{EqMeltGC}
\end{eqnarray}
\end{subequations}
which give the melting temperatures of GC and AT pairs in terms
of the (intermediate) salt-concentration  in the solvent obtained
from melting experiments \cite{kamanetskii71}. Note that the value
of $T_m^{AT}$ stems from an average over all possible combinations
of AT and TA base pairs \cite{krueger}; if only TA/AT and AT/TA
pairs are interchangeably used the value of $T_m^{AT}$ can be
lowered further. The above relations can be translated into
free-energy differences by $\Delta G=\Delta S(T_m-T)$, where $\Delta
S=-24.85$cal/(mol K) \cite{yakovchuk06}. Eqs.~\eqref{EqMeltAT},
\eqref{EqMeltGC} contain contributions from both base stacking and
hydrogen bonding, and is thus the melting temperature suitable for
our situation. It has been shown that the dependence on
salt-concentration lies in the stacking term \cite{yakovchuk06} and
not as previously thought in the hydrogen bonding term
\cite{protozanova04}. The stacking is a combination of hydrophobic,
electrostatic (screening of the negatively charged phosphate
groups), and dispersive interactions but there is no apparent
consensus on which term is the dominant one \cite{yakovchuk06}.  At
high salt-concentrations $\sim 1-5$M the temperature dependence
levels off due to a decrease in the hydrophobic effect; with most
water molecules tied up in the solvation of ions, the entropy
decrease involved in base stacking is small \cite{schildkraut65}.

The melting temperature of AT bonds has a stronger dependence on
salt-concentration,  so we can increase the ratio $u_{AT}/u_{GC}$ by
decreasing the salt-concentration. A further benefit is a lowering
of the melting temperatures thus enabling experiments well below
$T_m^{GC}\sim 100^o$C, which is the case when
$[\mathrm{Na}^+]=0.1$M, the standard concentration in
electrophoresis experiments. High temperatures have practical
disadvantages such as formation of air bubbles and increased
evaporation of solvent
molecules \cite{schildkraut65}.\\

Most relevant experiments on DNA have been conducted at $\sim 0.1$M
salt-concentration.  Those specifically looking at the salt
dependence of the melting temperature work in the range $\sim
0.01-1$M. A conservative estimate of $[\mathrm{Na}^+]=0.01$M gives
$u_{GC}\sim 1$ and $u_{AT}\simeq 6$ at $T\simeq 95^o$C, which is
sufficient for the Fokker-Planck approximation to be valid.

Concerning the possible length of a DNA construct it should be
reasonable to work  with segments up to $100-200$ bps. In a setup
combining fluorescence correlation spectroscopy and fluorescence
quenching, as introduced in Ref.~\cite{altan}, the DNA is free to
diffuse around in the solution. In this case the limiting factor is
the time it takes for the quencher to diffuse in and out of the
confocal volume.

Practically the experimental method of choice may be a dual optical
tweezers setup in which the DNA construct, via some handles of
double-stranded DNA, is connected to two beads held in place by the
tweezers. While this allows to keep the DNA construct in place the
force exerted on the chain is relatively small. However this would
allow direct observation of the construct avoiding diffusional
correction. The centre of the barrier region could be decorated with
either a fluorophore-quencher pair, or markers such as quantum dots
or small gold beads that can be visualized by a microscope. The
influence of the attached markers should decrease with longer
barrier length. Having this setup in a flow cell the system could be
triggered by flushing in a solution with either different
temperature or salt concentration. This can be done relatively
quickly \cite{gijs}. Once the two initial bubbles are thereby
created it should be possible to measure the coalescence time
calculated herein. Repeating the experiment would produce the
distribution of coalescence times, from which important system
parameters can be inferred. Decorating the barrier region with
several, sufficiently small, markers would, in principle, allow one
to measure the position of the coalescence.

\section{Final conclusions}\label{SecConclusions}

Single molecule techniques give us increasing insight into the
behavior, equilibrium and dynamic, of biopolymers. Of particular and
outstanding interest is DNA, due to its importance in biological
contexts as well as its role as a model biopolymer. In order to
extend our knowledge about the biological function of DNA it is
crucial to quantify and understand the denaturation behavior of DNA
at the single molecule level, its sequence dependence and,
ultimately, its relevance to genetic processes such as transcription
initiation. A major question hereby concerns the dynamics of
transient DNA denaturation bubbles.

While first single molecule fluorescence correlation experiments
have demonstrated the feasibility of monitoring the fluctuations of
a single bubble, some questions remain about the model system and
the explicit setup used in these experiments. In particular, the
obtained time scales for base pair zipping and unzipping as well as
the influence of the attached fluorophore-quencher pair remain under
debate.

Here we suggest an alternative model for accessing DNA stability
parameters and basepair (un-zipping) constants: in our model system
DNA bubbles in two AT-rich regions are formed and separated by a
more stable GC-rich barrier region. The coalescence behavior of the
two DNA bubbles across the barrier region is then studied. We show
that the stability parameters for bubble and barrier regions can
indeed be chosen sufficiently different to allow preparation of the
DNA construct in the proposed fashion, by the proper adjustment of
temperature and/or salt concentration. Once coalesced the newly
created single bubble is stabilized against immediate reclosure of
the barrier region both dynamically and due to the release of the
boundary free energy corresponding to one cooperativity factor
$\sigma_0$. Appropriate fluorophore-quencher tagging of the barrier
region basepairs should therefore allow for the direct observation
of the bubble merging dynamics.

Apart from the relevance of the investigated system for
understanding the dynamics of DNA and its biological function, the
mathematical description presented here is of interest for its own
sake as it corresponds to a previously unsolved case of two vicious
random walkers in opposite linear potentials. We established the
solution of this problem by solving a bivariate Fokker-Planck
equation (continuum limit of the discrete master equation
description) analytically. In a careful analysis we showed under
what conditions the Fokker-Planck approach is valid and what
deviations one would expect for realistic systems. Furthermore, the
analytic results were explained using qualitative arguments and
corroborated using stochastic simulations.

\acknowledgments The work of T.~N.\ is a part of the research plan
MSM 0021620834 financed by the Ministry of Education of the Czech
Republic and was also partly supported by the grant number
202/08/0361 of the Czech Science Foundation. T.~A.\ acknowledges
funding from the Knut and Alice Wallenberg foundation. R.~M.\
acknowledges the Natural Sciences and Engineering Research Council
(NSERC) of Canada, and the Canada Research Chairs programme, for
support. This work was started at CPiP 2005 (Computational Problems
in Physics, Helsinki, May 2005) supported by NordForsk, Nordita, and
Finnish NGSMP. We gratefully acknowledge very helpful discussions
with Oleg Krichevsky.

\begin{appendix}

\section{Calculation of $\tilde{p}(x;t|x_0)$ via Laplace transform}
\label{AppCalgt}

In this appendix we present a detailed calculation of the
single-walker auxiliary density $\widetilde{p}(x;t|x_0)$ satisfying
Eq.~\eqref{seq} together with the boundary conditions
Eqs.~(\ref{EqBoundxLgt}) and (\ref{EqBoundxRgt}), as well as the
initial condition $\widetilde{p}(x;t=0|x_0)=\delta(x-x_0)$.
$\widetilde{p}(x;t|x_0)$ solves the Schr{\"o}dinger equation
\begin{equation}
\frac{\partial}{\partial t} \widetilde{p}(x;t|x_0)=
\left[\frac{\partial^2}{\partial
x^2}-f^2\right]\widetilde{p}(x;t|x_0),
\end{equation}
that, after a Laplace transform and some rearrangement becomes
\begin{equation}
\label{EqLaplaceSchrod} \left[\frac{\partial^2}{\partial
x^2}-k(z)^2\right]\tilde{p}(x;z|x_0)= -\delta(x-x_0),
\end{equation}
with $k(z)=\sqrt{z+f^2}$ (we skip the explicit $z$-dependence in the
formulas from now on).

Consider first the solution of equation
\begin{equation}\label{gf}
\left[\frac{\partial^2}{\partial x^2}-k^2\right]g(x,x_0)=0,
\end{equation}
for $x<x_0$ with boundary condition (\ref{EqBoundxLgt}), and for
$x>x_0$ with boundary condition (\ref{EqBoundxRgt}). The solutions
are
\begin{equation}
g(x,x_0)=\left\{\begin{array}{ll} C_<\left(\kappa
e^{kx}+e^{-kx}\right), & x<x_0,\\[0.2cm]
C_>\left(e^{kx}+\kappa e^{2k}e^{-kx}\right), & x>x_0,
\end{array}
\right.
\end{equation}
with $\kappa\equiv (k+f)/(k-f)$.

The solution of Eq.~(\ref{EqLaplaceSchrod}) can now be found by the
ansatz
\begin{eqnarray}
\nonumber
&&\widetilde{p}(x;z|x_0)=\\
&&\left\{\begin{array}{ll} C \left(\kappa
e^{kx}+e^{-kx}\right)\left(e^{kx_0}+\kappa e^{2k}e^{-kx_0}
\right), & x<x_0,\\[0.2cm]
C\left(\kappa e^{kx_0}+e^{-kx_0}\right)\left(e^{kx}+\kappa
e^{2k}e^{-kx} \right), & x>x_0,
\end{array}\right.
\end{eqnarray}
such that
\begin{eqnarray}
\nonumber \widetilde{p}(x;z|x_0)
=C\left\{\kappa\left(e^{k(x+x_0)}+\kappa e^{2k}e^{-k|x-x_0|}\right)\right.\\
\left.+e^{k|x-x_0|}+\kappa e^{2k}e^{-k(x+x_0)}\right\},
\end{eqnarray}
where $C$ is determined from the jump condition by integrating
Eq.~(\ref{EqLaplaceSchrod})
\begin{eqnarray}
\nonumber
-1&=&\lim_{\epsilon\rightarrow0^+}\int_{x_0-\epsilon}^{x_0+\epsilon}
\left[\frac{\partial^2}{\partial x^2}-f^2\right]\widetilde{p}(x;z|x_0)dx\\
\nonumber &=&\lim_{\epsilon\rightarrow
0^+}\left\{\frac{\partial}{\partial
x}\widetilde{p}(x;z|x_0)\Big|_{x_0
+\epsilon}-\frac{\partial}{\partial
x}\widetilde{p}(x;z|x_0)\Big|_{x_0
-\epsilon}\right\}\\
&=&2kC\left(1-\kappa^2e^{2k}\right).
\end{eqnarray}
Here, it has been used that $\widetilde{p}(x;z|x_0)$ is continuous.

\section{Implementation of the discrete master equation}\label{AppMEnumerical}

\begin{figure}
  \begin{center}
\includegraphics[width=8cm]{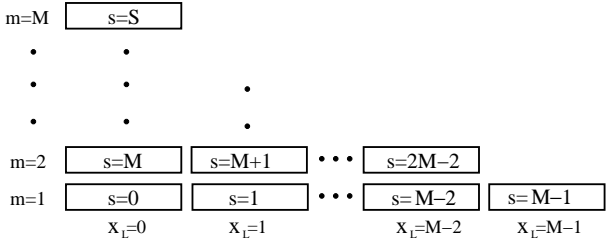}
  \end{center}
\caption{Enumeration scheme for the numerical analysis: The
two-dimensional grid points $(x_L,m)$ are replaced by a
one-dimensional running variable $s$. See text for details.}
\label{fig:s_space}
\end{figure}

To solve the eigenvalue equation (\ref{eq:eigenvalue_eq1}) by a
numerical scheme, it is convenient to replace the two-dimensional
grid points $(x_L,m)$ by a one-dimensional coordinate $s$ counting
all lattice points, compare with \cite{JPC}. We choose the
enumeration illustrated in figure \ref{fig:s_space}.
From this figure we notice that $m\in [1,M]$ and $x_L\in [0,M-m]$.
An arbitrary $s$-point can be obtained from a specific $(x_L,m)$
according to:
  \begin{equation}
s=(m-1)M-\frac{(m-1)(m-2)}{2}+x_L.\label{eq:s}
  \end{equation}
From this relation we notice that the maximum $s$ value is
  \begin{equation}
S={\rm max}\{ s\}=\frac{M(M+1)}{2}-1,
  \end{equation}
i.e., the size of the relevant $W$-matrix (see below) scales as
$M^2/2$.  Expression (\ref{eq:s}) allows us to change the transfer
coefficients to the $s$-variable, $t^\pm_{L/R} (x_L,m)\rightarrow
t^\pm_{L/R} (s)$, using the explicit expressions
(\ref{eq:t_L_plusB}), (\ref{eq:t_R_minusB}), (\ref{eq:t_L_minusB})
and (\ref{eq:t_R_plusB}) for the transfer coefficients, together
with the boundary conditions in equations (\ref{eq:reflecting2b})
and (\ref{eq:absorbingb}). From equation (\ref{eq:s}) and
Fig.~\ref{fig:s_space} we notice that [compare with Eq.
(\ref{eq:eigenvalue_eq1})]
  \begin{eqnarray}
s|_{x_L-1}^{m+1}&=&s|_{x_L}^m+M-m,\ {\rm for}\ x_L\ge 1,\nonumber\\
s|_{x_L+1}^{m-1}&=&s|_{x_L}^m-(M-m+1),\ {\rm for}\ m\ge 2,\nonumber\\
s|_{x_L}^{m-1}&=&s|_{x_L}^m-(M-m+2),\ {\rm for}\ m\ge 2,\nonumber\\
s|_{x_L}^{m+1}&=&s|_{x_L}^m+M-m+1,\nonumber\\
& & \ {\rm for}\ x_L\le M-(m+1) \ \& \ m\le M-1,\nonumber\\
  \end{eqnarray}
Eq. (\ref{eq:eigenvalue_eq1}) can then be written in matrix form
as
  \begin{equation}
\sum_{s'}W(s,s')Q_p(s')=-\eta_p Q_p(s),\label{eq:eigen2}
  \end{equation}
where explicitly the matrix-elements are
  \begin{eqnarray}
W(s,s+M-m)&=&t_L^+(s+M-m),\nonumber\\
&& {\rm for} \ s  \pitchfork x_L\ge 1,\nonumber\\
W(s,s-[M-m+1])&=&t_L^-(s-[M-m+1]),\nonumber\\
&&  {\rm for} \ s \pitchfork m\ge 2 \nonumber\\
W(s,s-[M-m+2])&=&t_R^+(s-[M-m+2]),\nonumber\\
&&  {\rm for} \ s \pitchfork m\ge 2, \nonumber\\
W(s,s+M-m+1)&=&t_R^-(s+M-m+1),\nonumber\\
&&  {\rm for} \ s \pitchfork  x_L\le M-(m+1) \nonumber\\
&& \& \ m\le M-1, \nonumber\\
W(s,s)&=&-(t_L^+(s)+t_L^-(s)\nonumber\\
&& +t_R^+(s)+t_R^-(s)),\label{eq:W}
  \end{eqnarray}
and the remaining matrix elements are equal to zero. We have
introduced the notation $s\pitchfork $ with the meaning ``$s$ is to
be taken for''. The problem at hand is that of determining the
eigenvalues and eigenvectors of the $(S+1)\times(S+1)$-matrix $W$
above. The coalescence time density is then calculated from Eqs.
(\ref{eq:fptd2}) and (\ref{eq:c_p}). In terms of the running
variable $s$, see Eq. (\ref{eq:s}), and the $W$-matrix defined in
equation (\ref{eq:W}) the detailed balance conditions
(\ref{eq:det_balance_1}) and (\ref{eq:det_balance_2}) become
\begin{equation}
\label{detail} W(s,s')\mathscr{Z}(s')=W(s',s)\mathscr{Z}(s).
\end{equation}
The orthogonality relation, Eq. (\ref{eq:ortho_relation}), becomes
\begin{equation}
\label{ortho} \sum_s\frac{Q_p(s)Q_{p'}(s)}{P_r^{\rm
eq}(s)}=\delta_{p,p'}.
\end{equation}
Convenient checks of the numerical results then include: (i) The
eigenvalues should be real and negative (so that $\eta_p>0$); (ii)
The eigenvectors should satisfy the orthonormality relation, Eq.
(\ref{ortho}).

\section{Stochastic simulation of bubble coalescence}\label{AppGillespie}

In this Section we give a brief introduction to the stochastic
simulation of DNA-breathing, for details we refer to
Ref.~\cite{suman}. We apply the Gillespie algorithm introduced in
$1976$ as a stochastic approach to the study of chemical reactions
\cite{gillespie76}.

Following the schematic of Fig.~\ref{fig:model}, we simulate the
dynamics of the two zipping forks separating the two initial
bubble domains from the barrier region. As each fork can either
zip or unzip, the system is described by the four different rates,
$t^{\mu}_{\nu}$, where $\mu\in\{+,-\}$, and
$\nu\in\{L,R\}$. Given these rates, we assume that the statistical
weight for a given event, $\{\mu,\nu\}$, to occur in a time
interval $[t,t+\delta t]$ is $t^{\mu}_{\nu}\delta t$.
Then the idea of the Gillespie scheme is the following
\cite{gillespie76}: The probability that nothing happens in the
time interval $[t,t+\tau]$, \emph{and\/} that in the following
interval $[t+\tau,t+\tau+d\tau]$ an event of type $\{\mu,\nu\}$
occurs, is the so-called \emph{reaction probability density\/}
\begin{equation}
\label{eq:jpd}
P(\tau,\mu,\nu)d\tau=P_0(\tau)t^{\mu}_{\nu}d\tau.
\end{equation}
To determine the probability $P_0(\tau)$ that no event happens
within $[t,t+\tau]$, this interval is divided into $K$ spans of
duration $\epsilon=\tau/K$. The probability that no event occurs
in the first subinterval $[t,t+\epsilon]$ is then
\begin{equation}
\prod_{\mu,\nu}\left[1-t^{\mu}_{\nu}\epsilon\right]=1-\sum_{\mu,\nu}
t^{\mu}_{\nu}\epsilon+\mathcal{O}(\epsilon^2).
\end{equation}
Treating the remaining intervals similarly produces an expression
for $P_0$,
\begin{eqnarray}
\nonumber
P_0(\tau)&=&\left[1-\sum_{\mu,\nu}t^{\mu}_{\nu}\epsilon+\mathcal{O}
(\epsilon^2)\right]^K\\
&=&\left[1-\sum_{\mu,\nu}t^{\mu}_{\nu}\tau/K+\mathcal{O}(K^{-2})
\right]^K.
\end{eqnarray}
Taking the limit $K\rightarrow\infty$ and reinserting in
Eq.~(\ref{eq:jpd}), we find the Poissonian law
\begin{equation}
\label{eq:Pdist}
P(\tau,\mu,\nu)=t^{\mu}_{\nu}\exp\left(-\sum_{\mu,\nu}t
^{\mu}_{\nu}\tau\right).
\end{equation}

At some given instant of time, $t$, the system is in a certain
configuration.
The update is performed as follows:\\
(i) The rates $t^{\mu}_{\nu}$ are calculated according to
the
configuration.\\
(ii) A set of random numbers $(\tau,\mu,\nu)$, distributed
according to
$P(\tau,\mu,\nu)$ in Eq.~(\ref{eq:Pdist}), is drawn from a generator.\\
(iii) The time is advanced according to $t\rightarrow t+\tau$, and
the
configuration is updated according to the randomly chosen event $\mu,\nu$.\\
The steps (i)-(iii) are repeated until a specified stop criterion
is fulfilled, in our case the merging of the two initial bubbles.
We record the stop time and the final configuration, and a new run
is initiated using the same initial condition.

Following Ref.~\cite{gillespie76} we briefly present how random
numbers $\tau$ and $\mu$ can be constructed using numbers drawn
from a uniform distribution: Let $P_c(\tau')$ be some continuous
probability density function, e.g., $P_c(\tau')d\tau$ is the
probability for finding a $\tau$ within the interval
$[\tau',\tau'+d\tau]$. The associated probability distribution
function is then defined as
\begin{equation}
F_c(\tau_0)=\int_{-\infty}^{\tau_0}P_c(\tau')d\tau',
\end{equation}
which is the probability of some $\tau$ being less than $\tau_0$.
To get a random $\tau$ according to $P_c$, given some random
number $R\in[0,1]$ drawn from the uniform distribution, we have to
invert $F_c(\tau)=R$. Using
$P_c(\tau)=\sum_{\mu,\nu}P(\tau,\mu,\nu)$ from
Eq.~(\ref{eq:Pdist}), with $\tau>0$ and inverting the expression,
we obtain
\begin{equation}
\tau=\frac{1}{\sum_{\mu,\nu}t^{\mu}_{\nu}}\ln\left(\frac{1}{R}\right).
\end{equation}
Similarly, we determine the appropriate random number for the direction of
the ``reaction'' (zipping/unzipping of left/right zipper fork), following
\begin{equation}
F_d(\mu_0)=\sum_{\nu=1}^{\mu_0}P_d(\nu),
\end{equation}
is the probability of having $\mu\leq\mu_0$. Inversion given some
random number $R\in[0,1]$ drawn from the uniform distribution is
now requiring that $F_d(\mu-1)<R<F_d(\mu)$. Using $P_d(\mu)=\int
P(\tau,\mu)d\tau$ the random event $\mu$ is determined by
\begin{equation}
    \sum_{\nu=1}^{\mu-1}r_{\nu}<R\sum_{\nu=1}^Nr_{\nu}\leq \sum_{\nu=1}^{\mu}r_{\nu}.
\end{equation}

\end{appendix}


\begin{thebibliography}{99}

\bibitem{watson} J.~D.~Watson and F.~H.~C.~Crick, \emph{Nature} \textbf{171},
737 (1953).

\bibitem{cantor} C.~R.~Cantor and P.~R.~Schimmel, \emph{Biophysical
Chemistry} (W.~H.~Freeman, New York, 1980).

\bibitem{kornberg} A.~Kornberg, \emph{DNA synthesis} (W.~H.~Freeman,
San Francisco, 1974).

\bibitem{kornberg1} A.~Kornberg and T.~A.~Baker, \emph{DNA Replication}
(W.~H.~Freeman, New York, 1992).

\bibitem{frank} M.~D.~Frank-Kamenetskii, \emph{Phys.\ Rep.}
\textbf{288}, 13 (1997).

\bibitem{delcourt} S.~G.~Delcourt and R.~D.~Blake, \emph{J.\ Biol.\ Chem.} \textbf{266},
15160 (1991).

\bibitem{santalucia} R.~D.~Blake, J.~W.~Bizzaro, J.~D.~Blake, G.~R.~Day,
S.~G.~Delcourt, J.~Knowles, K.~A.~Marx, and J.~SantaLucia, Jr.,
\emph{Bioinf.} \textbf{15}, 370 (1999).

\bibitem{krueger} A.~Krueger, E.~Protozanova, and M.~D.~Frank-
Kamenetskii, \emph{Biophys.\ J.} \textbf{90}, 3091 (2006).

\bibitem{poland} D.~Poland and H.~A.~Scheraga,\emph{Theory of helix-coil transitions in biopolymers}
  (Academic Press, New York, 1970).

\bibitem{peyrard_np} M.~Peyrard, \emph{Nature Phys.} \textbf{2}, 13 (2006).

\bibitem{wartell} R.~M.~Wartell and A.~S.~Benight, \emph{Phys.\ Rep.} \textbf{126}, 67
(1985).

\bibitem{guttmann} C.~Richard and A.~J.~Guttmann, \emph{J.\ Stat.\ Phys.} \textbf{115}, 925
(2004).

\bibitem{yeramian} E. Yeramian, \emph{Gene} {\bf 255}, 139 (2000);
\emph{ibid.} 151 (2000).

\bibitem{carlon} E. Carlon, M. L. Malki, and R. Blossey,
\emph{Phys.\ Rev.\ Lett.} {\bf 94}, 178101 (2005).

\bibitem{gueron} M.~Gu{\'e}ron, M.~Kochoyan, and J.-L.~Leroy, \emph{Nature} \textbf{328}, 89 (1987).

\bibitem{altan} G.~Altan-Bonnet, A.~Libchaber, and O.~Krichevsky, \emph{Phys.\ Rev.\ Lett.} \textbf{90}, 138101
(2003).

\bibitem{ctn} R.~Metzler, T.~Ambj\"{o}rnsson, A.~Hanke, Y.~Zhang, and S.~Levene, \emph{J.\ Comput.\ Theor.\ Nanoscience} \textbf{4}, 1 (2007).

\bibitem{pant} K.~Pant, R.~L.~Karpel, and M.~C.~Williams, \emph{J.\
Mol.\ Biol.} \textbf{327}, 571 (2003).

\bibitem{pant1} K.~Pant, R.~L.~Karpel, I.~Rouzina, and M.~C.~Williams,
\emph{J.\ Mol.\ Biol.} \textbf{336}, 851 (2004); \emph{ibid.}
\textbf{349}, 317 (2005).

\bibitem{somepanwill} I.~M.~Sokolov, R.~Metzler, K.~Pant, and M.~C.~Williams,
\emph{Biophys.\ J.} \textbf{89}, 895 (2005).

\bibitem{tobiasrc}
T.~Ambj{\"o}rnsson and R.~Metzler, \emph{Phys.\ Rev.\ E}
\textbf{72}, 030901(R) (2005).

\bibitem{choi} C.~H.~Choi, G.~Kalosakas, K.~{\O}.~Rasmussen, M.~Hiromura,
A.~R.~Bishop, and A.~Usheva, \emph{Nucleic Acids Res.} \textbf{32},
1584 (2004).

\bibitem{kalosakas} S.~Ares and G.~Kalosakas, \emph{Nano Lett.} \textbf{7} (2), 307 (2007).

\bibitem{tobiasprl} T.~Ambj\"{o}rnsson, S.~K.~Banik, O.~Krichevsky, and
R.~Metzler, \emph{Phys.\ Rev.\ Lett.} \textbf{97}, 128105 (2006).

\bibitem{tobiasbj} T.~Ambj\"{o}rnsson, S.~K.~Banik, O.~Krichevsky, and
R.~Metzler, \emph{Biophys.\ J.} \textbf{92}, 2674 (2007).

\bibitem{peyrard} M. Peyrard and A. R. Bishop, \emph{Phys.\ Rev.\ Lett.} \textbf{62},
2755 (1989).

\bibitem{dauxois} T. Dauxois, M. Peyrard, and A. R. Bishop, \emph{Phys.\
Rev.\ E} \textbf{47}, R44 (1993).

\bibitem{peyrard_cm} B.~S.~Alexandrov, L.~T.~Wille, K.~{\O}.~Rasmussen, A.~R.~Bishop, and K.~B.~Blagoev,
\emph{Phys.\ Rev.\ E} \textbf{74}, 050901 (2006).

\bibitem{campa} A.~Campa and A.~Giansanti, \emph{Phys.\ Rev.\ E} \textbf{58}, 3585
(1998).

\bibitem{peyrard_nonl} M.~Peyrard, \emph{Nonlinearity} \textbf{17},
R1 (2004).

\bibitem{hame} A.~Hanke and R.~Metzler, \emph{J.\ Phys.\ A} \textbf{36},
L473 (2003).

\bibitem{kafri} A.~Bar, Y.~Kafri, and D.~Mukamel \emph{Phys.\ Rev.\ Lett.} \textbf{98},
038103 (2007).

\bibitem{hans} H.~C.~Fogedby and R.~Metzler, \emph{Phys.\ Rev.\ Lett.}
\textbf{98}, 070601 (2007); \emph{Phys. Rev. E} \textbf{76}, 061915 (2007).

\bibitem{bicout} D.~J.~Bicout and E.~Kats, \emph{Phys.\ Rev.\ E} \textbf{70}, 010902(R)
(2004).

\bibitem{JPC}
T.~Ambj{\"o}rnsson and R.~Metzler, \emph{J.\ Phys:\ Cond.\ Matt.}
\textbf{17}, S1841 (2005).

\bibitem{suman} S.~K.~Banik, T.~Ambj{\"o}rnsson, and R.~Metzler, \emph{Europhys.\ Lett.} \textbf{71}, 852 (2005).

\bibitem{hwa} T.~Hwa, E.~Marinari, K.~Sneppen, and L.-H.~Tang, \emph{Proc.\ Natl.\
Acad.\ Sci.\ USA} \textbf{100}, 4411 (2003).

\bibitem{jeon} J.-H.~Jeon, P.~J.~Park, and W.~Sung, \emph{J.\ Chem.\ Phys.} \textbf{125}, 164901
(2006).

\bibitem{tobiaspre} T.~Ambj\"{o}rnsson, S.~K.~Banik, M.~A.~Lomholt, and
R.~Metzler, \emph{Phys.\ Rev.\ E} \textbf{75}, 021908 (2007).

\bibitem{tobiasctn} R.~Metzler and T.~Ambj\"{o}rnsson, \emph{J.\ Comp.\
Theoret.\ Nanosc.} \textbf{2}, 389 (2005).

\bibitem{tobiasjpc} T.~Ambj\"{o}rnsson and R.~Metzler, \emph{J.\ Phys:\ Cond.\
Matt.} \textbf{17}, S4305 (2005).

\bibitem{poland1} D.~Poland and H.~A.~Scheraga, \emph{J.\ Chem.\ Phys.} \textbf{45}, 1464 (1966).

\bibitem{fisher1} M.~E.~Fisher, \emph{J.\ Chem.\ Phys.} \textbf{45}, 1469 (1966).

\bibitem{kafri1} Y.~Kafri, D.~Mukamel, and L.~Peliti, \emph{Phys.\ Rev.\ Lett.} \textbf{85},
4988 (2000).

\bibitem{comment} A.~Hanke and R.~Metzler, \emph{Phys.\ Rev.\ Lett.}
\textbf{90}, 159801 (2003); Y.~Kafri, D.~Mukamel, and L.~Peliti, \emph{ibid.}
159802 (2003).

\bibitem{haochme} A. Hanke, M. G. Ochoa, and R. Metzler, \emph{Phys. Rev.
Lett.} \textbf{100}, 018106 (2008).

\bibitem{blossey} R.~Blossey and E.~Carlon, \emph{Phys.\ Rev.\ E} \textbf{68}, 061911
(2003).

\bibitem{beacon} G.~Bonnet, O.~Krichevsky, and A.~Libchaber, \emph{Proc.\ Natl.\
Acad.\ Sci.\ USA} \textbf{95}, 8602 (1998); G.~Bonnet, S.~Tyagi,
A.~Libchaber, and F.~R.~Kramer, \emph{Proc.\ Natl.\ Acad.\ Sci.\
USA} \textbf{96}, 6171 (1999).

\bibitem{olegrev} O.~Krichevsky and G.~Bonnet, \emph{Rep.\ Prog.\ Phys.} \textbf{65}, 251
(2002).

\bibitem{fisher} M.~E.~Fisher, \emph{J.\ Stat.\ Phys.} \textbf{34}, 667 (1984).

\bibitem{bray} A.~J.~Bray and K.~Winkler, \emph{J.\ Phys.\ A} \textbf{37}, 5493 (2004).

\bibitem{fixman} M.~Fixman and J.~J.~Freire, \emph{Biopol.} {\bf 16},
2693 (1977).

\bibitem{Di_Marzio_Guttman_Hoffman}
E.~A.~Di Marzio, C.~M.~Guttman, and J.~D.~Hoffman, \emph{Faraday
Discuss.} {\bf68}, 210 (1979).

\bibitem{twobubb} T.~Novotn\'{y}, J.~N.~Pedersen, M.~S.~Hansen,
T.~Ambj{\"o}rnsson, and R.~Metzler, \emph{Europhys.\ Lett.}
\textbf{77}, 48001 (2007).

\bibitem{Risken} H.~Risken, \emph{The Fokker-Planck Equation}
(Springer, Berlin, 1989).

\bibitem{van_Kampen} N.~G.~van~Kampen, \emph{Stochastic Processes in Physics and Chemistry} (North-Holland, Amsterdam, 2nd
  ed., 1992).

\bibitem{Gardiner}
C.~W.~Gardiner, \emph{Handbook of Stochastic Methods for Physics,
Chemistry and the Natural Sciences} (Springer, Berlin, 1989).

\bibitem{pawula} R. F. Pawula, Phys. Rev. \textbf{162}, 186 (1967).

\bibitem{marcienkiewicz} J. Marcinkiewicz, Math. Z. \textbf{44}, 612 (1939).


\bibitem{redner} S.~Redner, \emph{A Guide to First-Passage
Processes} (Cambridge University Press, Cambridge UK, 2001).

\bibitem{Novotny2000} T.~Novotn\'{y} and P.~Chvosta, \emph{Phys.\ Rev.\ E} \textbf{63}, 012102 (2000). %; cond-mat/0004351v2.

\bibitem{nome} T.~Novotn\'{y} and R.~Metzler, in preparation (2008).

\bibitem{aslangul} C. Aslangul, J. Phys. A 32, 3993 (1999).

\bibitem{tobiasbob} T.~Ambj\"ornsson and R.~J.~Silbey. J. Chem. Phys. 129,
165103 (2008).

\bibitem{kamanetskii71} M. D. Frank-Kamenetskii, \emph{Biopol.},
            {\bf 10}, 2623 (1971).

\bibitem{yakovchuk06} P.~Yakovchuk, E.~Protozanova and M. D. Frank-Kamenetskii,
    \emph{Nuc. Acid. Res.} {\bf 34}, 564 (2006).

\bibitem{protozanova04} E.~Protozanova, P.~Yakovchuk and M.~D.~Frank-Kamenetskii,
    \emph{ J. Mol. Biol.} {\bf 342}, 775 (2004).

\bibitem{schildkraut65} C.~Schildkraut and S.~Lifson,
        \emph{ Biopol.} {\bf 3}, 195 (1965).

\bibitem{gijs} B. van den Broek, M. A. Lomholt, S.-M. J. Kalisch, R. Metzler,
and G. J. L. Wuite, Proc. Natl. Acad. Sci. USA \textbf{105}, 15738
(2008).

\bibitem{gillespie76} D.~T.~Gillespie, \emph{Jour.\ Comp.\ Phys.} \textbf{22}, 403
(1976).

\end{thebibliography}
\end{document}